\documentclass[useAMS,usenatbib]{mn2e}
\usepackage{times}
\usepackage{psfig}
\usepackage{lscape}

\newcommand{\mpass}{~mag~arcsec$^{-2}$}
\newcommand{\bmgc}{B_{\mbox{\tiny \sc MGC}}}
\newcommand{\eref}[1]{(\ref{#1})}
\newcommand{\dsc}[1]{_{\mbox{\tiny \sc #1}}}
\newcommand{\usc}[1]{^{\mbox{\tiny \sc #1}}}
\renewcommand{\d}{{\rm d}}
\newcommand{\ion}[2]{#1$\,${\sc #2}}

\title[MGC compact galaxies]{The Millennium Galaxy Catalogue: a census
of local compact galaxies}
\author[J.~Liske et~al.]{J.~Liske,$^1$\thanks{E-mail: jliske@eso.org}
S.~P.~Driver,$^2$ P.~D.~Allen,$^2$ N.~J.~G.~Cross$^3$ and
R.~De~Propris$^{4,5}$\\
$^1$European Southern Observatory, Karl-Schwarzschild-Str.~2, 85748
Garching, Germany\\
$^2$Research School of Astronomy \& Astrophysics, Australian National
University, Cotter Road, Weston, ACT 2611, Australia\\
$^3$Institute for Astronomy, University of Edinburgh, Royal
Observatory, Blackford Hill, Edinburgh EH9 3HJ\\
$^4$H.H.~Wills Physics Laboratory, University of Bristol, Tyndall
Avenue, Bristol BS8 1TL\\
$^5$Cerro Tololo Inter-American Observatory, Casilla 603, La Serena,
Chile\\}

\begin{document}

\date{Accepted
...... Received .....}

\pagerange{\pageref{firstpage}--\pageref{lastpage}} \pubyear{2006}

\maketitle

\label{firstpage}

\begin{abstract}
We use the Millennium Galaxy Catalogue (MGC) to study the effect of
compact galaxies on the local field galaxy luminosity function. Here
we observationally define as `compact' galaxies that are too small to
be {\em reliably} distinguished from stars using a standard
star-galaxy separation technique. In particular, we estimate the
fraction of galaxies that are misclassified as stars due to their
compactness.

We have spectroscopically identified {\em all} objects to $\bmgc =
20$~mag in a $1.14$~deg$^2$ sub-region of the MGC, regardless of
morphology. From these data we develop a model of the high surface
brightness incompleteness and estimate that $\sim$$1$ per cent of
galaxies with $\bmgc < 20$~mag are misclassified as stars, with an
upper limit of $2.3$ per cent at $95$ per cent confidence. However,
since the missing galaxies are preferentially sub-$L^*$ their effect
on the faint end of the luminosity function is substantially
amplified: we find that they contribute $\sim$$6$ per cent to the
total luminosity function in the range $-17 < M_B < -14$~mag, which
raises the faint end slope $\alpha$ by $0.03^{+0.02}_{-0.01}$. Their
contribution to the total $B$-band luminosity density is $\sim$$2$ per
cent.  Roughly half of the missing galaxies have already been
recovered through spectroscopy of morphologically stellar targets
selected mainly by colour. We find that the missing galaxies mostly
consist of intrinsically small, blue, star-forming, sub-$L^*$ objects.

In combination with the recent results of \citet{Driver05} we have now
demonstrated that the MGC is free from both high and low surface
brightness selection bias for giant galaxies ($M_B \la
-17$~mag). Dwarf galaxies, on the other hand, are significantly
affected by these selection effects. To gain a complete view of the
dwarf population will require both deeper and higher resolution
surveys.
\end{abstract}


\section{Introduction}
\label{intro}
The idea that galaxies at a given luminosity span a wide range in
surface brightness (SB) and that selection effects might cause both
high and low SB galaxies to be under-represented in the observed
galaxy distribution goes back to at least
\citet{Reaves56,Zwicky57,Arp65} and \citet{Disney76}. Low SB galaxies
are difficult to detect because of their low contrast against the sky
\citep{Impey97} while high SB galaxies may be confused with stars.
Previously overlooked types of galaxies have indeed been uncovered
over time, such as the giant low SB galaxy Malin~1 \citep{Bothun87}
and several types of blue compact galaxies (\citealp{Zwicky66}; see
\citealp{Kunth00} for a recent review) at various luminosities and
redshifts, including \ion{H}{ii} galaxies
\citep[e.g.][]{Sargent70,Telles97}, blue compact dwarfs
\citep[e.g.][]{Thuan81,Cairos01} and luminous compact blue galaxies
\citep{Phillips97,Guzman03} which in turn include the compact narrow
emission line galaxies (CNELGs) of
\citet{Koo94,Koo95,Guzman96,Guzman98}. Most recently a potentially new
class of faint and compact objects, the so-called ultra compact dwarfs
(UCDs), was discovered in the Fornax
(\citealp{Hilker99,Drinkwater00a,Drinkwater03}; see also
\citealp*{Mieske04a}), Abell 1689 \citep{Mieske04b} and Virgo clusters
\citep{Hacsegan05}.

The extent to which a given survey misses these sorts of galaxies
depends on the survey's detailed selection limits in the
luminosity--SB plane (LSP). These in turn are determined by the
survey's apparent magnitude, SB and size limits, i.e.\ by its depth
and resolution.  Recently, \citet{Driver05} presented the bivariate
brightness distribution of the Millennium Galaxy Catalogue
\citep[MGC;][]{Liske03b}, i.e.\ the space density of galaxies in the
LSP. They used simulations to determine the MGC's selection limits and
found that the observed galaxy distribution was well separated from
both the low and high SB selection limits for luminosities brighter
than at least $M_B \approx -18$~mag (see their fig.\ 12). Hence SB
selection effects are not a significant issue in the MGC at these
luminosities (unless the SB distribution of galaxies is bimodal) and
in that sense the bright end of MGC's luminosity function (LF) is
truly `global'. The goal is now to obtain a similarly complete picture
at lower luminosities.

To probe the galaxies beyond the MGC's {\em low} SB limit requires
deeper imaging and spectroscopy on 8-m class telescopes.

In this paper we will instead address the question of what lies beyond
the MGC's {\em high} SB selection limit. In particular, what is the
effect of the compact galaxies beyond this limit on the local field
galaxy LF and what is their contribution to the luminosity density? To
answer these questions we must first understand and quantify the MGC's
incompleteness due to some compact galaxies having been misclassified
as stars from the imaging data. To this end we have assembled two
datasets:

First, we conducted an all-object spectroscopic survey in the range
$16 \le \bmgc < 20$~mag over a $1.14$~deg$^2$ sub-region of the MGC,
targeting {\em all} detected objects in this magnitude range,
regardless of classification, morphology, colour or any other
properties. This complete sample will be used to determine the
fraction of galaxies that have been misclassified due to their compact
nature.

A few such all-object surveys have been performed previously:
\citet*{Morton85} observed all $606$ objects to $B \le 20$~mag in a
$0.31$~deg$^2$ region while \citet{Colless90,Colless91} investigated a
random sample of $266$ objects in the range $21 \le b_{\rm J} \le
23.5$~mag. Most recently, \citet{Drinkwater00b} described an ambitious
project to observe $14\,000$ objects with $16.5 \le b_{\rm J} \le
19.7$~mag in a $12$~deg$^2$ region centered on the Fornax cluster,
which led to the discovery of $13$ compact galaxies behind the Fornax
cluster (mostly CNELG-like objects but also including four redder,
weaker emission line galaxies; \citealt{Drinkwater99}) as well as the
UCDs in the cluster itself mentioned above.

Unfortunately, these previous studies cannot be used to reliably and
quantitatively predict the number of compact galaxies misclassified as
stars by the MGC because they either lack SB measurements or were
performed in a special environment. On the other hand, many studies
concerning the known compact galaxy populations contain very detailed
SB information but they do not use complete samples, and hence cannot
be used to make precise predictions either.

The second dataset consists of all available spectroscopy of stellar
objects in the range $16 \le \bmgc < 20$~mag in the full MGC survey
region. We have augmented existing public data (almost exclusively of
QSO candidates) with observations of additional stellar objects.
Although these data are incomplete they provide a lower limit on the
frequency of misclassified compact galaxies and help in the
investigation of their properties.


The paper is organised as follows: we describe our data in Section
\ref{data}, study the properties of compact galaxies masquerading as
stars in Section \ref{mcprops} and quantify the MGC's incompleteness
due to misclassification in Section \ref{mgc_inc}. In Section \ref{cglf}
we evaluate the contribution of compact galaxies to the local galaxy
luminosity function and density. Finally, we summarise our findings in
Section \ref{conclusions}. We use $H_0 = 100 \,
h$~km~s$^{-1}$~Mpc$^{-1}$, $\Omega_{\rm M} = 0.3$ and $\Omega_\Lambda
= 0.7$ throughout.

\section{Data}
\label{data}

\subsection{The MGC}
\label{mgcdata}
The Millennium Galaxy Catalogue \citep[MGC;][]{Liske03b} is a deep
($\mu_{\mbox{\tiny \sc lim}} = 26$\mpass, $B_{\mbox{\tiny \sc lim}} =
24$~mag), wide-field ($37.5$~deg$^2$) $B$-band imaging survey
conducted with the Wide Field Camera on the $2.5$-m Isaac Newton
Telescope. The survey region is a $72$-deg long, $35$-arcmin wide
strip along the equator, and is fully contained within the survey
regions of the Sloan Digital Sky Survey \citep[SDSS;][]{York00} and
the 2dF Galaxy Redshift Survey \citep[2dFGRS;][]{Colless01}.

Details of the observations, reduction, object detection using {\sc
SExtractor} \citep{Bertin96} and catalogue construction are given by
\citet{Liske03b}. As it pertains to the present investigation we
briefly remind the reader of our star/galaxy separation procedure for
objects with $16 \le \bmgc < 20$~mag.

As a starting point for classification we used the stellaricity
parameter provided by {\sc SExtractor}, which is produced for each
object by an artificial neural network. Its input consists of nine
object parameters (eight isophotal areas and the peak intensity) and
the seeing. The output consists of a single number, called
stellaricity, which takes a value of $1$ for unresolved objects, $0$
for extended objects and intermediate values in more dubious cases.
We began by classifying all objects with stellaricity $\ge 0.98$ as
stellar. This is a `natural' value to adopt because the stellaricity
distribution rises sharply from $0.97$ to $0.98$ (see fig.~9 of
\citealp{Liske03b}). All remaining objects were then inspected visually
and classified as stars, galaxies, asteroids (defined as apparently
real objects without any detectable counterparts in SuperCOSMOS Sky
Survey or SDSS images), or detections due to cosmic rays, satellite
trails, CCD defects, etc.

If an object was found to be incorrectly deblended or if the object's
parameters were obviously wrong, the object was re-extracted by
manually changing the {\sc SExtractor} extraction parameters until a
satisfactory result was achieved. In addition, all low-quality regions
in the survey (e.g.\ near CCD edges or defects) were carefully masked
out, resulting in an effective survey area of $30.88$~deg$^2$.

\subsection{Spectroscopy}
To obtain spectroscopy for the MGC ($\bmgc < 20$~mag) we first turned
to publicly available data and matched the MGC to the SDSS Data
Release 1 \citep[SDSS-DR1;][]{Abazajian03}, 2dFGRS, 2QZ \citep{Croom04}
and other smaller surveys. We then conducted our own redshift survey
(MGCz), mainly using the Two Degree Field (2dF) facility on the
Anglo-Australian Telescope (see \citealp{Driver05} for details). While
the primary goal of this campaign was to obtain redshifts for those
MGC galaxies without public data, we also targeted two other samples:

\subsubsection{The all-object sub-region}
\label{aoregion}
The first of these additional samples consisted of {\em all} objects,
irrespective of morphology, image classification or colour, in a small
sub-region of the MGC. This all-object sub-region is defined as that
part of the MGC strip which is bounded by $11^{\rm h} 49^{\rm m}
52\fs44 < \alpha < 12^{\rm h} 00^{\rm m} 08\fs45$ (J2000), i.e.\ MGC
fields 56--61 (with only partial coverage of the end fields). The
seeing in these fields ranged from $1.05$ to $1.25$~arcsec with the
median at $1.15$~arcsec, slightly better than the survey median of
$1.27$~arcsec. The effective area of this region, after subtraction of
the exclusion regions, is $1.1371$~deg$^2$ and it contains $1552$
objects (and $10$ asteroids) in the range $16 \le \bmgc < 20$~mag. The
bright limit is set by the magnitude where stars begin to flood and
the faint limit is that of the main MGCz galaxy survey.

\begin{figure}
\psfig{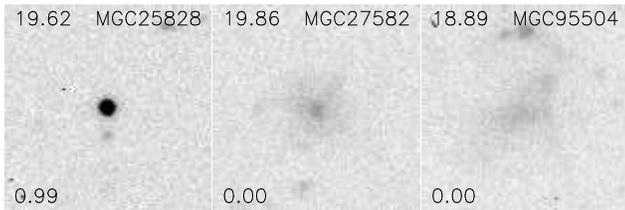}
\caption{MGC postage stamps of the three objects in the all-object
  region without a good quality ($Q_z \ge 3$) spectrum. The labels
  give the object ID (upper right), $\bmgc$ (upper left) and
  stellaricity (lower left) for each object. The image sizes are $33
  \times 33$~arcsec$^2$.}
\label{nospec}
\end{figure}

We have at least one spectrum for each object in this sample and we
give details of the spectroscopic identifications in Table
\ref{aotab}. However, for three objects the quality of the data is too
low to allow an identification or redshift measurement ($Q_z = 2$, see
\citealp{Driver05}). Hence the overall spectroscopic completeness is
$99.8$ per cent. The three objects without good quality spectra are
shown in Fig.~\ref{nospec}. For two of these (MGC27582 and MGC95504)
there is no possibility of misclassification: they are clearly low SB
galaxies and we will treat them as confirmed galaxies in the rest of
the paper.

In contrast, the third object, MGC25828 (J$115340.51$ $+$$001407.9$),
is unresolved (stellaricity $= 0.99$) and hence could be a
misclassified galaxy. Since the number of such objects is small (cf.\
Table \ref{aotab}) even a single additional case would make a
significant difference to our incompleteness estimates. Hence we take
a closer look at this object in the following.

\begin{figure}
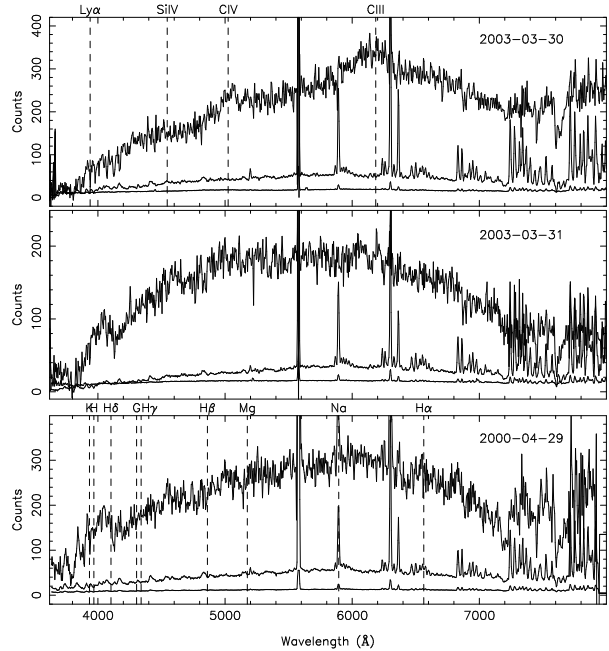

\psfig{file=mgc_cg_fig2a.ps,angle=-90,width=\columnwidth}
\psfig{file=mgc_cg_fig2b.ps,angle=-90,width=\columnwidth}
\psfig{file=mgc_cg_fig2c.ps,angle=-90,width=\columnwidth}
\caption{Three independent 2dF spectra of the possible blazar
  MGC25828. The top two spectra are from MGCz, while the bottom
  spectrum was obtained by the 2QZ. None of them are
  flux-calibrated. Each panel shows the object, mean sky and error
  spectra. The purpose of the sky spectra is to indicate the positions
  of strong sky lines and they have been scaled arbitrarily. The UT
  date of observation is also shown in each panel. The vertical dashed
  lines indicate the positions of prominent broad QSO emission lines
  at $z=2.24$ (top panel) and stellar absorption features at $z=0$
  (bottom panel).}
\label{blazar}
\end{figure}

We have three independent 2dF spectra for MGC25828, one of which was
contributed by the 2QZ (who could not identify it either), and these
are shown in Fig.~\ref{blazar}. Even though a high signal-to-noise
continuum is common to all three spectra (S/N $\approx 12$) we were
nevertheless unable to confidently identify this object because of the
lack of emission and absorption features. A broad `bump' near
$4040$~\AA\ is only evident in the bottom two spectra while the top
spectrum shows two broad bumps at $5025$ and $6185$~\AA. If this
variation is real, i.e.\ due to object variability and not some
instrumental effect, then the object must vary on the timescale of one
day (time difference between the top two spectra), while the
similarity between the bottom two spectra (which were taken three
years apart) must then be coincidental.

A strong optical continuum with no or only weak broad emission lines,
and variability on the timescale of days are characteristics of
blazars \citep[e.g.][]{Wagner95}. Indeed, if we interpret the broad
bumps above as weak emission lines then they could possibly be
identified as Ly$\alpha$, \ion{C}{iv} and \ion{C}{iii}] at $z \approx
2.24$ (cf.\ Fig.~\ref{blazar}).

The possible identification of MGC25828 as a blazar is further
supported by two facts: (i) The object has AGN-like colours as
evidenced by its selection as a target by both the 2QZ and the SDSS
QSO survey (who have not yet observed it). (ii) According to the
SuperCOSMOS Sky Survey \citep{Hambly01c} and USNO-B \citep{Monet03}
catalogues this object has no significant proper motion. Hence it is
not surprising that MGC25828 satisfies the \citet{Londish02} selection
criteria for BL Lac candidates in the 2QZ (even though it is missing
from their published list of candidates; Londish, priv.\ comm.).

\begin{table}
\caption{Spectroscopic identifications of the complete sample of
objects in the all-object region ($16 \le \bmgc < 20$~mag).}
\label{aotab}
\begin{center}
\begin{minipage}{4.9cm}
\begin{tabular}{lr}
\hline
Description & Number\\
\hline 
Galaxies & $444$\\
MCs$^a$ & $3$\\
Stars & $1083$\\ 
QSOs & $21$\\ 
Unidentified (blazar or WD) & $1$\\ 
\hline
Total & $1552$\\
\hline
\end{tabular}\\
$^a$Spectroscopically identified galaxies which had been
morphologically misclassified as stars.\\
\end{minipage}
\end{center}
\end{table}

On the other hand we do not observe any broad-band photometric
variability. Comparing the object's brightness in the SDSS and MGC
images, which were taken one year apart, we find a difference of only
$0.07$~mag (see \citealp{Cross04b} for the filter conversion). Also,
the object does not appear to be a strong radio or X-ray source since
it is not listed in either the FIRST or {\it ROSAT} All-Sky Survey
(RASS) catalogues.

The lack of stellar absorption features in the spectrum of MGC25828
(cf.\ bottom panel of Fig.~\ref{blazar}) means that this object could
also be a DC white dwarf \citep[e.g.][]{Wesemael93}. However, it also
implies that the observed continuum is unlikely to be due to a stellar
population and so we can discard the possibility that this object is a
galaxy. We are therefore reasonably certain to have identified {\em
all} galaxies in the all-object region (within our magnitude limits).

\subsubsection{Additional spectroscopy of stellar objects}
\label{morestel}

\begin{figure}
\psfig{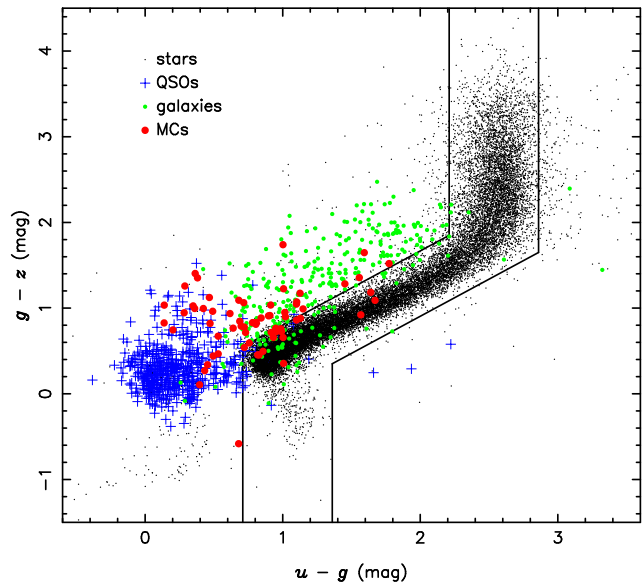}
\caption{Colour--colour plot using SDSS-DR1 PSF magnitudes showing our
  colour selection of targets for additional observations of
  morphologically stellar objects outside of the all-object
  region. Black points and blue crosses show MGC stars and QSOs with
  $18 \le \bmgc < 20$~mag respectively. We define the band between the
  solid lines as the `stellar locus' in this colour--colour
  space. Green dots show a random sample of $300$ galaxies in the same
  magnitude range. Note that PSF magnitudes are physically not very
  meaningful for extended objects; we only wish to indicate the degree
  of separation between stars and galaxies in this space. Red dots
  show all $65$ spectroscopically identified galaxies in the MGC which
  had been morphologically misclassified as stars (MCs).}
\label{col_sel}
\end{figure}

Apart from the main galaxy survey and the all-object region, we also
observed a third sample of targets when 2dF fibres were available.
The target list for this sample comprised: (i) all morphologically
stellar objects with $18 \le \bmgc < 20$~mag that lie away from the
stellar locus in $(u-g)$--$(g-z)$ colour--colour space, where the
`stellar locus' is defined in Fig.~\ref{col_sel}; (ii) a random subset
of stellar objects that lie in the stellar locus ($18 \le \bmgc <
20$~mag); (iii) stellar objects morphologically classified as galaxies
by the SDSS-DR1 ($16 \le \bmgc < 20$~mag). Objects were only targeted
if they had not been observed previously by other surveys. Merging the
incomplete observations of this target list with publicly available
data yields $3223$ secure identifications of morphologically stellar
objects outside of the all-object region, $65$ per cent of which were
supplied by MGCz. Details are given in Table \ref{astab}. The
relatively high discovery rate of misclassified galaxies compared to
the previous Section is presumably due to the colour
selection. However, this sample is in no way complete. Since more than
one survey has significantly contributed to this sample and since the
magnitude limits, colour selection techniques and incompleteness
levels of these surveys are different, it is very difficult to
reliably construct its selection function. For example, we expect the
spectroscopic completeness to the left of the stellar band in
Fig.~\ref{col_sel} to be very different from that within the band
because of the MGCz colour selection. However, even the space to the
left will not be sampled homogeneously because of the colour selection
of the 2QZ and SDSS QSO surveys. Simply ignoring these additional data
does not help either because we specifically excluded objects already
observed by these surveys from MGCz observations.

In Section \ref{incmc} we will ignore these complications and apply a
rudimentary incompleteness correction to this sample based on the
selection function of the MGCz observations. In any case, at the very
least this sample provides an indication of the properties of
misclassified compact galaxies as well as a lower limit on their
incidence.

\begin{table}
\caption{Additional spectroscopic identifications of morphologically
stellar objects outside of the all-object region ($16 \le \bmgc <
20$~mag).}
\label{astab}
\begin{center}
\begin{minipage}{3.5cm}
\begin{tabular}{lr}
\hline
Description & Number \\
\hline 
MCs$^a$ & $62$\\
Stars & $2625$\\ 
QSOs & $536$\\ 
\hline
Total & $3223$\\
\hline
\end{tabular}\\
$^a$As in Table \ref{aotab}.
\end{minipage}
\end{center}
\end{table}

\section{Properties of Misclassified Compact Galaxies}
\label{mcprops}

\begin{figure*}
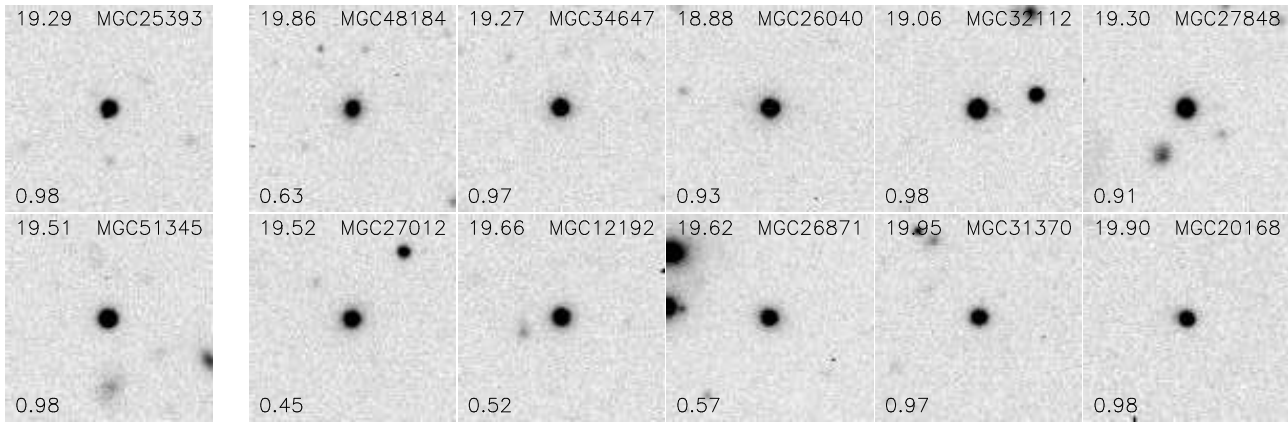

\begin{minipage}{\textwidth}
\psfig{file=mgc_cg_fig4a.ps,angle=-90,height=2.8cm}
\end{minipage}
\hspace{3mm}
\begin{minipage}{\textwidth}
\psfig{file=mgc_cg_fig4b.ps,angle=-90,height=2.8cm}
\end{minipage}\\
\vspace{-0.5mm}
\begin{minipage}{\textwidth}
\psfig{file=mgc_cg_fig4c.ps,angle=-90,height=2.8cm}
\end{minipage}
\hspace{3mm}
\begin{minipage}{\textwidth}
\psfig{file=mgc_cg_fig4d.ps,angle=-90,height=2.8cm}
\end{minipage}
\caption{Right: MGC postage stamps of a sample of ten misclassified
  compact galaxies (MCs). Left: Images of two stars for comparison.
  Labels and image sizes are as in Fig.~\ref{nospec}.}
\label{mcimg}
\end{figure*}

\begin{figure}
\psfig{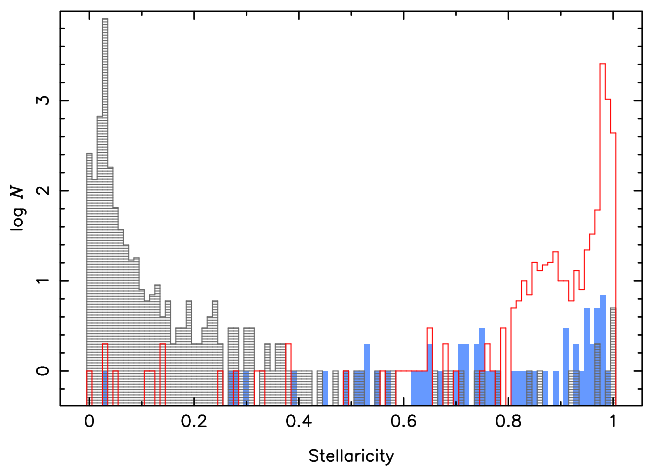}
\caption{Stellaricity distributions of spectroscopically confirmed
  stars and QSOs (open red histogram), galaxies (hashed) and
  misclassified compact galaxies (MCs, solid blue).}
\label{stell}
\end{figure}

In total we have spectroscopically identified $65$ compact galaxies
(cf.\ Tables \ref{aotab} and \ref{astab}) that had been
morphologically misclassified as stars from the imaging data (using
the procedure described in Section \ref{mgcdata}). In the following we
will refer to these objects as misclassified compact galaxies (MCs).
We illustrate their morphological similarity to stars in
Fig.~\ref{mcimg}, where we show a selection of their MGC images
alongside the images of two stars. In Fig.~\ref{stell} we compare
their stellaricity values to the stellaricity distributions of stellar
objects and non-misclassified galaxies.

In this Section we will investigate the properties of these $65$ MCs
in order to understand what types of galaxies are being missed due to
misclassification. However, since these $65$ MCs make up only about
one half of the total number of MCs expected in the MGC (see Section
\ref{mgc_inc}) it is in principle possible that their properties are
not fully representative of all MCs in general. Nevertheless, they
provide a good indication of the types of objects that may be missed.

\subsection{Spectral properties}
\label{specprops}
\begin{figure}
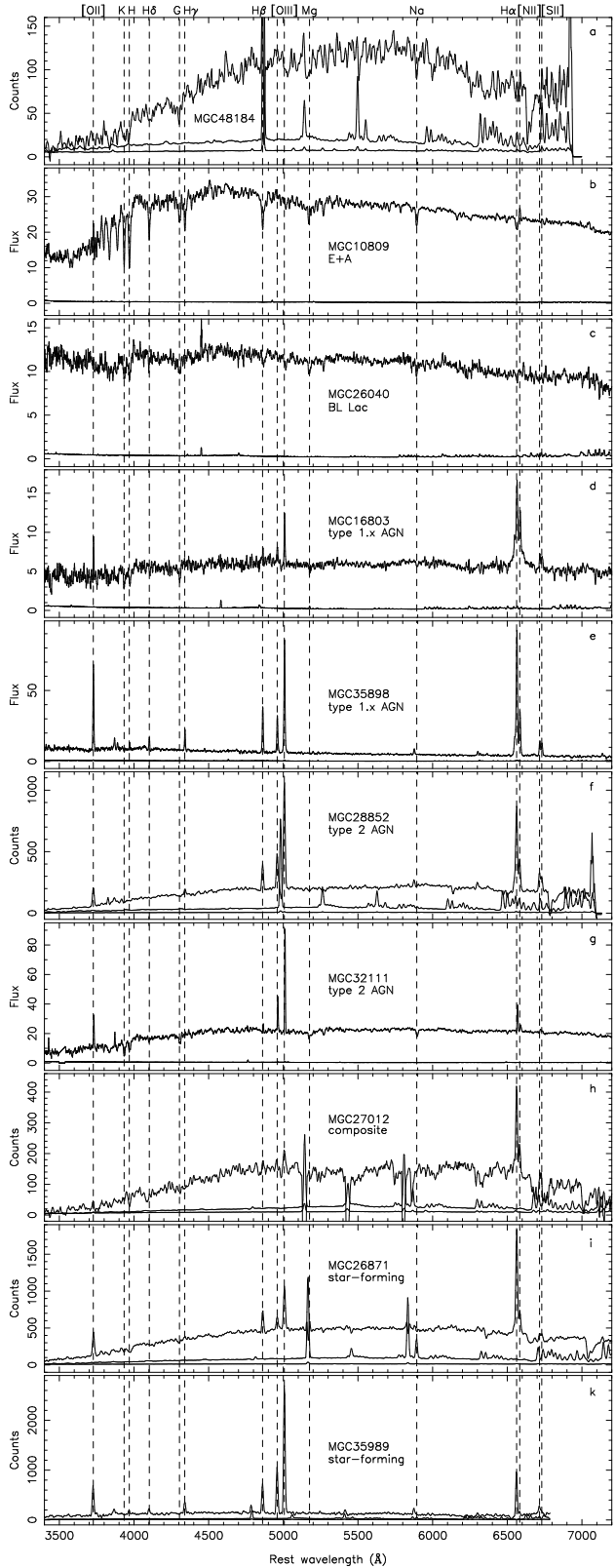

\psfig{file=mgc_cg_fig6a.ps,angle=-90,width=\columnwidth}
\psfig{file=mgc_cg_fig6b.ps,angle=-90,width=\columnwidth}
\psfig{file=mgc_cg_fig6c.ps,angle=-90,width=\columnwidth}
\psfig{file=mgc_cg_fig6d.ps,angle=-90,width=\columnwidth}
\psfig{file=mgc_cg_fig6e.ps,angle=-90,width=\columnwidth}
\psfig{file=mgc_cg_fig6f.ps,angle=-90,width=\columnwidth}
\psfig{file=mgc_cg_fig6g.ps,angle=-90,width=\columnwidth}
\psfig{file=mgc_cg_fig6h.ps,angle=-90,width=\columnwidth}
\psfig{file=mgc_cg_fig6i.ps,angle=-90,width=\columnwidth}
\psfig{file=mgc_cg_fig6k.ps,angle=-90,width=\columnwidth}
\vspace{-2mm}
\caption{Rest-frame spectra of ten MCs representative of the various
  spectal classifications. We show the object, error and mean sky
  spectra (where available) which have been scaled arbitrarily. Only
  SDSS spectra are flux calibrated. The vertical dashed lines mark the
  locations of common nebular emission and stellar absorption
  features. (a) Absorption line galaxy. (b) Example E+A galaxy. (c) BL
  Lac. (d)--(e) Example MCs requiring a broad emission line component
  at least in H$\alpha$. (f)--(k) Example MCs classified as type 2
  AGN, composite or star-forming galaxies on the basis of
  Fig.~\ref{bpt}.}
\label{mcspec}
\end{figure}

Only five of the $65$ MCs are absorption line galaxies and we show
three example spectra in Fig.~\ref{mcspec}(a)--(c). Three of the
absorption line MCs (MGC10809, MGC17413 and MGC34647) appear to be E+A
galaxies \citep{Dressler83} since they have strong Balmer absorption
with an H$\delta$ rest equivalent width (EW) of $>5$~\AA\ superimposed
on an elliptical galaxy spectrum. Although MGC10809 possibly has a
weak [\ion{O}{ii}] $\lambda 3727$ emission line (cf.\
Fig.~\ref{mcspec}b), its EW is $< 2.5$~\AA\ and hence all three
objects satisfy commonly used criteria to define E+A galaxies
\citep[e.g.][]{Goto05}.

Note also the unusual spectrum of MGC26040 (Fig.~\ref{mcspec}c). It
has a very weak $4000$~\AA\ break ($D_{4000} = 1.28$) and a blue,
AGN-like $(u-g)$ colour (cf.\ Section \ref{colprops}). It is also a
radio and X-ray source, with detections in FIRST, NVSS and RASS. Based
on these properties \citet{Schwope00} and \citet{Anderson03}
independently classified MGC26040 as a probable BL Lac. This object
also erroneously appears in the SDSS QSO catalogue of
\citet{Schneider03}. Erroneously because \citeauthor{Schneider03}
required a spectrum to show at least one broad emission line for the
object to be included in the catalogue, and stressed that BL Lac
objects were not included. Indeed, the object has been excised from
the most recent version of the SDSS QSO catalogue \citep{Schneider05}.

The spectra of $60$ of our $65$ MCs show nebular emission lines. The
EUV photons that ionize the emitting gas may be provided by either hot
massive stars from recent or ongoing star formation or by the
non-thermal continuum of an active galactic nucleus (AGN) or both. In
principle, the emission line spectrum is quite sensitive to the
hardness of the ionizing radiation and hence emission line ratio
diagrams are frequently used to distinguish between star-forming or
\ion{H}{ii} region-like galaxies and narrow-lined (i.e.\ type 2) AGN,
such as Seyfert 2 galaxies and low-ionization nuclear emission regions
(LINERs). These line ratio diagnostics were pioneered by
\citet*{Baldwin81} and further refined by \citet{Veilleux87}.

To construct an [\ion{O}{iii}]/H$\beta$ versus [\ion{N}{ii}]/H$\alpha$
line ratio diagram we have measured the line fluxes of the $60$
emission line MCs by first subtracting a locally fitted continuum and
then fitting one or more gaussians as required. The
H$\alpha$-[\ion{N}{ii}] complex was fitted with at least three
components where the [\ion{N}{ii}] $\lambda\lambda$6548,6584 doublet
was constrained to have the same width and redshift and a flux ratio
of 1:3. For two galaxies we were able to improve the fit by adding a
weak broad component in H$\alpha$ and H$\beta$ while five galaxies
required a weak broad component in H$\alpha$ only, indicative of
Seyfert $1.8$ and $1.9$ galaxies respectively. Hence we classify these
seven objects as AGN and collectively refer to them as type 1.x
AGN. Two example spectra are shown in Fig.~\ref{mcspec}(d)--(e).

\begin{figure}
\psfig{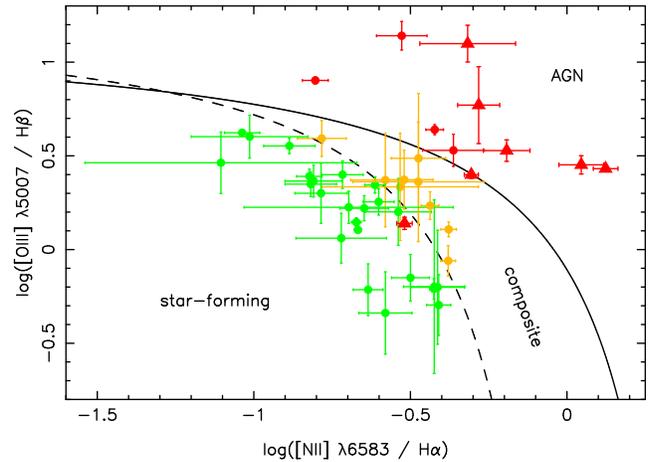}
\caption{Emission line ratio diagnostic diagram used to classify
  emission line MCs. The dashed line shows the observed upper boundary
  of star-forming galaxies \citep{Kauffmann03a} while the solid line
  gives the theoretical upper limit for starbursts
  \citep{Kewley01}. Accordingly, green dots mark star-forming
  galaxies, red dots denote type 2 AGN and orange dots show
  composites. Red triangles show the type 1.x AGN, using only their
  narrow line components.}
\label{bpt}
\end{figure}

In Fig.~\ref{bpt} we show the resulting line ratios for $44$ of the
$60$ emission line MCs. The other $16$ MCs are excluded from this plot
because either their spectra do not cover H$\alpha$ ($11$ cases) or at
least one of the four lines could not be detected with reasonable S/N
($4$ cases). Finally, one object was excluded because both the
H$\alpha$ and [\ion{O}{iii}] $\lambda 5007$ lines appeared to be
saturated.

The dashed line in Fig.~\ref{bpt} is the empirical upper envelope of
star-forming galaxies found by \citet{Kauffmann03a} and the solid line
shows the theoretical upper limit for the location of pure
starbursting galaxies in this diagram derived by \citet{Kewley01}. A
number of galaxies lie between these lines and it is clear that it is
not possible to perfectly separate star-forming galaxies from type 2
AGN. The reason is of course that nuclear activity and star-formation
are not mutually exclusive phenomena and an object's emission lines
may contain contributions from both. It is well known that
star-formation may affect the line ratios not only of the lower
luminosity LINERs \citep[e.g.][]{Ho93} but also of the more powerful
Seyfert 2s \citep[e.g.][]{CidFernandes01}.

Hence we follow \citet{Brinchmann04} and divide our MCs into three
groups: we label objects below the dashed line in Fig.~\ref{bpt} as
star-forming galaxies ($24$ objects, green dots), those above the
solid line as type 2 AGN ($4$ objects, red dots) and those in-between
as composites ($9$ objects, orange dots). In Fig.~\ref{mcspec}(f)--(k)
we show example type 2, composite and star-forming spectra.

We caution the reader that the above terminology is somewhat
misleading because it suggests that all star-formation/AGN overlap is
limited to the composite objects. That is not the case: we plot in
Fig.~\ref{bpt} as red triangles the narrow components of the seven
galaxies already identified as AGN by the presence of broad
components. One of these objects lies below the
\citeauthor{Kauffmann03a} line, demonstrating that even objects
labelled as `star-forming' above may in fact contain an AGN \citep[see
also][]{Simpson05}. Similarly, a significant fraction of bona fide
Seyfert 2s are known to contain a young stellar population
\citep[e.g.][]{StorchiBergmann00} and \citet{Brinchmann04} estimated
that nearly $5$ per cent of the total star formation rate density at
$z \sim 0.1$ occurs in the host galaxies of type 2 AGN.

In addition, two further caveats apply to our classifications based on
our emission line measurements: (i) Our 2dF spectra (from 2dFGRS and
MGCz) are not flux calibrated. However, since the members of each line
pair are quite close in wavelength we do not expect the flux
calibration to have a large impact on the line ratios. (ii) We do not
account for the underlying stellar Balmer absorption. Neglecting this
absorption pushes galaxies in Fig.~\ref{bpt} toward the AGN regime. To
assess the severity of this problem we compare our line ratio
measurements to those of \citet{Tremonti04} who carefully modelled and
subtracted the stellar continuum (including the Balmer absorption)
before emission line fitting. Eight of our objects in Fig.~\ref{bpt}
were also in \citeauthor{Tremonti04}'s sample, but six of the $16$
line ratio measurements cannot be compared to ours because of our use
of a broad component. For the rest we find that
\citeauthor{Tremonti04}'s ratios are smaller than ours by less than
$0.05$ dex in five cases, by $0.07$ dex in one case, and by
$\sim$$0.2$ dex in four cases. From Fig.~\ref{bpt} we can see that a
shift of the latter magnitude would change the classification of three
of the type 2 objects and of most of the composites.

Of the $11$ MCs whose spectra do not cover H$\alpha$ we find that six
show evidence of a broad component in H$\beta$ and we add these six
objects to the type 1.x AGN class. One further object (MGC21061) has
an [\ion{O}{iii}]/H$\beta$ ratio of $1.4$, placing it firmly in the
AGN regime of Fig.~\ref{bpt} irrespective of its unknown
[\ion{N}{ii}]/H$\alpha$ ratio. In fact this object shows broad
\ion{Mg}{ii} emission and so we also classify it as a type 1.x AGN.

This leaves a total of nine emission line objects which we are unable
to classify because at least one pair of emission lines is unavailable
for various reasons. However, two of these objects appear to be almost
exact superpositions of a small, faint galaxy and a foreground star,
since they both show clear evidence of stellar absorption lines at
$z=0$ in addition to redshifted emission lines.

\begin{landscape}
\begin{table}
\begin{minipage}{20cm}
\caption{Properties of all known $65$ MCs in the MGC.}
\label{mcpropstab}
\begin{tabular}{cccrrrrrcl}
\hline
ID & $z$ & $M_{\bmgc}^a$ & $R_e^b$ & $(u-g)_0^c$ & $(g-z)_0^c$
& $\log \frac{\mbox{[\ion{O}{iii}]}}{{\rm H}\beta}$ & $\log \frac{\mbox{[\ion{N}{ii}]}}{{\rm H}\alpha}^d$
& Class$^e$ & Comments$^f$\\
\hline
\multicolumn{3}{l}{Absorption line objects:}\\
MGC48184 J$133616.13$$+$$000726.3$ & $0.1461$ & $-19.04$ & $1.24$ & $ 1.81$ & $ 1.27$ &         &         & 		& \\
MGC10809 J$104329.26$$+$$001205.1$ & $0.1326$ & $-20.18$ & $1.07$ & $ 1.72$ & $ 0.95$ &         &         & E+A		& DR1: gal\\
MGC17413 J$111146.63$$+$$000017.9$ & $0.1104$ & $-19.34$ & $0.99$ & $ 1.66$ & $ 1.05$ &         &         & E+A		& DR1: gal\\
MGC34647 J$123243.45$$-$$000453.8$ & $0.1875$ & $-19.98$ & $0.75$ & $ 1.77$ & $ 0.70$ &         &         & E+A		& \\
MGC26040 J$115404.55$$-$$001009.8$ & $0.2535$ & $-20.76$ & $1.05$ & $ 0.35$ & $ 1.03$ &         &         & BL Lac	& Sw, A: BL Lac; RASS; FIRST; DR1: gal; in all-object region\\
\\
\multicolumn{3}{l}{Emission line objects:}\\
MGC00704 J$100140.13$$-$$000123.2$ & $0.3497$ & $-20.72$ & $1.93$ & $ 0.49$ & $ 0.96$ & $ 0.53$ & $-0.19$ & t1.x AGN	& bc H$\alpha$; S; DR1: gal\\
MGC03855 J$101326.04$$-$$000136.4$ & $0.2555$ & $-20.13$ & $1.84$ & $ 1.00$ & $ 1.74$ & $ 1.18$ &   ---   & t1.x AGN	& bc H$\beta$; RASS; FIRST\\
MGC16803 J$111011.90$$+$$000352.7$ & $0.2177$ & $-19.37$ & $1.63$ & $ 0.29$ & $ 1.26$ & $ 0.77$ & $-0.28$ & t1.x AGN	& bc H$\alpha$; A: broad-line AGN; RASS; DR1: QSO\\
MGC20403 J$112525.59$$-$$000301.5$ & $0.3596$ & $-20.83$ & $1.87$ & $ 0.28$ & $ 0.95$ & $ 0.67$ &   ---   & t1.x AGN	& bc H$\beta$\\
MGC21061 J$112753.52$$+$$000518.0$ & $0.4596$ & $-21.75$ & $2.96$ & $ 0.36$ & $ 1.00$ & $ 1.36$ &   ---   & t1.x AGN	& bc \ion{Mg}{ii}; S; FIRST; DR1: gal\\
MGC29999 J$121040.15$$-$$000315.5$ & $0.3016$ & $-20.17$ & $2.63$ & $ 0.36$ & $ 1.40$ &         &   ---   & t1.x AGN?	& bad sky near H$\beta$; bc H$\beta$?\\
MGC32112 J$122102.95$$-$$000733.3$ & $0.3662$ & $-21.46$ & $1.55$ & $ 0.62$ & $ 0.94$ & $ 0.14$ & $-0.52$ & t1.x AGN	& bc H$\alpha$,H$\beta$; S; A,W: NLSy1; RASS; FIRST; DR1: gal; 2QZ: NELG\\
MGC33386 J$122804.66$$-$$000506.7$ & $0.3190$ & $-20.78$ &  --- & $ 0.14$ & $ 0.83$ & $ 0.36$ &   ---   & t1.x AGN	& bc H$\beta$; 2QZ: QSO\\
MGC35898 J$123840.24$$-$$000744.0$ & $0.2046$ & $-19.64$ & $1.87$ & $ 0.45$ & $ 0.34$ & $ 0.43$ & $ 0.12$ & t1.x AGN	& bc H$\alpha$; K: LINER; FIRST; DR1: gal\\
MGC41402 J$130611.20$$-$$000932.7$ & $0.3925$ & $-20.92$ & $2.40$ & $ 0.38$ & $ 1.35$ & $ 0.45$ & $ 0.05$ & t1.x AGN	& bc H$\alpha$, \ion{Mg}{ii}; S; DV: variable; DR1: QSO\\
MGC55396 J$140434.62$$-$$000800.1$ & $0.3034$ & $-20.30$ & $2.55$ & $ 0.42$ & $ 0.99$ & $ 0.40$ & $-0.31$ & t1.x AGN	& bc H$\alpha$; K: LINER; DR1: gal; 2QZ: NELG\\
MGC57877 J$141207.92$$+$$000325.3$ & $0.2344$ & $-20.05$ & $1.39$ & $ 0.47$ & $ 0.82$ & $ 1.04$ &   ---   & t1.x AGN	& bc H$\beta$\\
MGC61874 J$142631.76$$-$$000249.8$ & $0.2928$ & $-20.19$ & $2.22$ & $ 0.14$ & $ 1.04$ &         &   ---   & t1.x AGN	& bad sky near [\ion{O}{iii}]; bc H$\beta$; 2QZ: NELG\\
MGC67450 J$144413.26$$+$$001118.4$ & $0.2067$ & $-20.11$ & $2.45$ & $ 0.47$ & $ 1.12$ & $ 1.10$ & $-0.32$ & t1.x AGN	& bc H$\alpha$,H$\beta$\\
MGC27848 J$120118.84$$-$$000654.6$ & $0.1818$ & $-19.86$ & $1.36$ & $ 1.03$ & $ 0.61$ & $ 0.53$ & $-0.36$ & t2 AGN	& \\
MGC28852 J$120450.47$$-$$001159.5$ & $0.1196$ & $-18.46$ & $0.92$ & $ 0.74$ & $ 0.72$ & $ 0.64$ & $-0.42$ & t2 AGN	& \\
MGC32111 J$122217.54$$-$$000744.1$ & $0.1714$ & $-20.26$ & $2.50$ & $ 1.79$ & $ 1.16$ & $ 1.14$ & $-0.53$ & t2 AGN	& Ka; FIRST; DR1: gal\\
MGC63901 J$143053.30$$-$$000917.5$ & $0.1231$ & $-19.89$ & $1.77$ & $ 0.87$ & $ 0.64$ & $ 0.90$ & $-0.80$ & t2 AGN	& FIRST\\
MGC02455 J$100702.72$$-$$000937.2$ & $0.0666$ & $-17.34$ & $0.93$ & $ 1.07$ & $ 0.90$ & $ 0.59$ & $-0.78$ & C		& \\
MGC20976 J$112754.40$$-$$001341.7$ & $0.0491$ & $-16.28$ & $0.42$ & $ 1.03$ & $ 0.71$ & $ 0.33$ & $-0.53$ & C		& \\
MGC21348 J$113005.55$$-$$001001.7$ & $0.0684$ & $-17.07$ & $0.68$ & $ 0.93$ & $ 0.88$ & $ 0.37$ & $-0.58$ & C		& \\
MGC22397 J$113626.84$$+$$000456.7$ & $0.1338$ & $-18.66$ & $0.86$ & $ 0.90$ & $ 0.52$ & $-0.06$ & $-0.38$ & C		& 2QZ: NELG\\
MGC27012 J$115741.37$$-$$001332.7$ & $0.0851$ & $-17.76$ & $0.84$ & $ 1.16$ & $ 0.87$ & $ 0.49$ & $-0.47$ & C		& in all-object region\\
MGC34229 J$123121.18$$-$$000647.2$ & $0.0793$ & $-17.10$ & $0.73$ & $ 1.14$ & $ 0.65$ & $ 0.36$ & $-0.47$ & C		& \\
MGC54685 J$140108.42$$-$$000030.4$ & $0.1117$ & $-18.13$ & $0.93$ & $ 1.06$ & $ 0.76$ & $ 0.23$ & $-0.44$ & C		& \\
MGC56265 J$140726.91$$+$$001154.1$ & $0.1756$ & $-19.92$ & $1.60$ & $ 1.20$ & $ 0.74$ & $ 0.11$ & $-0.38$ & C		& \\
MGC96267 J$111523.85$$+$$000622.5$ & $0.0277$ & $-14.65$ & $0.46$ & $ 0.64$ & $ 0.81$ & $ 0.37$ & $-0.52$ & C		& \\
MGC04637 J$101532.73$$-$$001337.6$ & $0.0943$ & $-18.10$ & $1.40$ & $ 1.30$ & $ 1.03$ & $-0.34$ & $-0.58$ & SF		& \\
MGC12192 J$104947.47$$+$$001423.1$ & $0.1215$ & $-18.56$ & $0.99$ & $ 1.21$ & $ 0.93$ & $-0.20$ & $-0.41$ & SF		& \\
MGC14297 J$105917.28$$+$$001232.8$ & $0.1993$ & $-20.24$ & $2.32$ & $ 1.22$ & $ 0.63$ & $-0.15$ & $-0.50$ & SF		& \\
MGC15183 J$110403.13$$+$$000644.0$ & $0.0760$ & $-17.74$ & $0.89$ & $ 0.80$ & $ 0.51$ & $ 0.40$ & $-0.72$ & SF		& \\
MGC16641 J$111014.20$$-$$000022.9$ & $0.0919$ & $-17.89$ & $0.81$ & $ 0.63$ & $ 0.50$ & $ 0.11$ & $-0.67$ & SF		& DR1: gal\\
MGC22174 J$113423.54$$-$$001444.1$ & $0.0770$ & $-17.28$ & $0.61$ & $ 0.82$ & $ 0.78$ & $ 0.39$ & $-0.82$ & SF		& \\
MGC23590 J$114130.16$$-$$000433.5$ & $0.1107$ & $-17.90$ & $1.27$ & $ 0.82$ & $ 0.56$ & $ 0.22$ & $-0.65$ & SF		& \\
MGC26871 J$115821.52$$-$$000520.6$ & $0.0799$ & $-17.51$ & $0.77$ & $ 1.06$ & $ 0.77$ & $ 0.34$ & $-0.61$ & SF		& RASS; in all-object region\\
MGC27394 J$120013.37$$-$$000747.7$ & $0.0614$ & $-16.56$ & $0.35$ & $ 0.86$ & $ 0.50$ & $ 0.60$ & $-1.01$ & SF		& \\
MGC27446 J$120016.76$$-$$001119.0$ & $0.0807$ & $-17.81$ & $0.92$ & $ 1.05$ & $ 0.75$ & $ 0.20$ & $-0.54$ & SF		& \\
\hline
\end{tabular}\\
\end{minipage}
\end{table}
\end{landscape}

\newpage

\begin{landscape}
\begin{table}
\begin{minipage}{20cm}
\contcaption{} 
\begin{tabular}{cccrrrrrcl}
\hline
ID & $z$ & $M_{\bmgc}^a$ & $R_e^b$ & $(u-g)_0^c$ & $(g-z)_0^c$
& $\log \frac{\mbox{[\ion{O}{iii}]}}{{\rm H}\beta}$ & $\log \frac{\mbox{[\ion{N}{ii}]}}{{\rm H}\alpha}^d$
& Class$^e$ & Comments$^f$\\
\hline
MGC27813 J$120141.90$$-$$000502.8$ & $0.0829$ & $-17.69$ & $0.51$ & $ 1.07$ & $ 0.71$ & $-0.20$ & $-0.42$ & SF		& \\
MGC31360 J$121830.69$$-$$001117.2$ & $0.0849$ & $-18.07$ & $0.68$ & $ 0.89$ & $ 0.58$ & $ 0.15$ & $-0.67$ & SF		& DR1: gal\\
MGC31370 J$121754.42$$-$$001147.8$ & $0.0708$ & $-16.87$ & $0.44$ & $ 0.76$ & $ 0.83$ & $-0.30$ & $-0.41$ & SF		& \\
MGC33857 J$123007.02$$-$$001123.4$ & $0.0905$ & $-17.62$ & $0.64$ & $ 1.19$ & $ 1.03$ & $-0.21$ & $-0.64$ & SF		& \\
MGC35989 J$124032.96$$-$$001402.7$ & $0.1646$ & $-19.25$ & $1.35$ & $ 0.54$ & $ 0.04$ & $ 0.62$ & $-1.04$ & SF		& 2QZ: NELG\\
MGC47144 J$133229.12$$+$$001425.2$ & $0.0804$ & $-17.18$ & $0.40$ & $ 0.94$ & $ 0.55$ & $ 0.23$ & $-0.70$ & SF		& \\
MGC52270 J$135230.99$$+$$000345.7$ & $0.0833$ & $-17.55$ & $0.67$ & $ 1.00$ & $ 0.69$ & $ 0.30$ & $-0.79$ & SF		& 2QZ: NELG\\
MGC57241 J$140846.88$$-$$001243.5$ & $0.0950$ & $-17.73$ & $0.91$ & $ 0.91$ & $ 0.49$ & $ 0.46$ & $-1.11$ & SF		& \\
MGC60772 J$142229.09$$-$$000812.6$ & $0.0543$ & $-16.21$ & $0.45$ & $ 0.84$ & $ 0.75$ & $ 0.35$ & $-0.82$ & SF		& \\
MGC64370 J$143243.74$$-$$000105.8$ & $0.1303$ & $-18.48$ & $0.88$ & $ 0.77$ & $ 0.35$ & $ 0.37$ & $-0.81$ & SF		& \\
MGC65997 J$143754.03$$-$$001527.0$ & $0.1400$ & $-18.51$ & $1.24$ & $ 0.78$ & $ 0.52$ & $ 0.06$ & $-0.72$ & SF		& \\
MGC67179 J$144048.09$$-$$000858.1$ & $0.1212$ & $-18.61$ & $1.01$ & $ 0.57$ & $ 0.21$ & $ 0.55$ & $-0.89$ & SF		& \\
MGC95276 J$105526.00$$-$$000501.8$ & $0.1081$ & $-19.03$ & $1.57$ & $ 1.14$ & $ 1.09$ & $ 0.25$ & $-0.60$ & SF		& \\
MGC96202 J$101416.19$$+$$001133.4$ & $0.0713$ & $-18.40$ & $0.87$ & $ 1.20$ & $ 0.92$ & $-0.21$ & $-0.42$ & SF		& \\
MGC00145 J$095931.46$$-$$000013.3$ & $0.2152$ & $-19.64$ & $1.33$ & $ 1.01$ & $ 0.59$ & $ 0.16$ &   ---   & 		& \\
MGC34325 J$123100.44$$+$$000402.0$ & $0.4633$ & $-21.14$ & $2.61$ & $ 0.85$ & $ 0.20$ & $ 0.36$ &   ---   & t1.x AGN?	& 2QZ: NELG\\
MGC11178 J$104353.73$$-$$001419.0$ & $0.1005$ & $-18.41$ & $1.65$ & $ 1.24$ & $ 0.91$ &         &         & 		& weak emission lines\\
MGC14790 J$110156.38$$-$$000001.8$ & $0.0733$ & $-17.03$ & $0.53$ & $ 1.50$ & $ 1.20$ &         &         & 		& weak emission lines\\
MGC17184 J$111122.09$$-$$000636.9$ & $0.1231$ & $-18.36$ & $1.12$ & $ 1.26$ & $ 0.82$ &         & $-0.46$ & 		& weak emission lines\\
MGC35480 J$123642.47$$-$$000946.6$ & $0.0722$ & $-16.97$ & $0.43$ & $ 0.72$ & $-0.50$ &         &         & 		& H$\alpha$,[\ion{O}{iii}] saturated; 2QZ: NELG\\
MGC52589 J$135243.21$$-$$000017.3$ & $0.1589$ & $-19.64$ & $1.12$ & $ 0.91$ & $ 0.65$ & $-0.14$ &         & 		& H$\alpha$ in telluric line\\
MGC20168 J$112428.81$$+$$000321.7$ & $0.3283$ & $-20.71$ & $1.26$ & $ 2.19$ & $ 0.56$ & $ 0.50$ &   ---   & star+gal	& \\
MGC28037 J$120107.34$$-$$001639.4$ & $0.2056$ & $-20.50$ & $1.23$ & $ 1.25$ & $ 0.14$ &         &   ---   & star+gal	& weak H$\beta$\\
\hline
\end{tabular}\\
$^a$Absolute magnitude using $h_{100} = 1$, $\Omega_{\rm M} = 0.3$,
$\Omega_\Lambda = 0.7$. For the BL Lac and type 1.x objects we assume
a power-law spectral shape and hence a $K$-correction of $K(z) =
-(\alpha + 1) \, 2.5 \log(1+z)$, where $\alpha = -0.5$. For all other
objects we use the $K$-corrections of \citet[][section 3.2]{Driver05}.\\
$^b$Physical half-light radius in units of $h^{-1}$ kpc, where the
observed apparent half-light radius, $r_e$, has been seeing-corrected
using $r_0^2 = r_e^2 - 0.32 \, \Gamma^2$ \citep[][Appendix A]{Driver05},
where $\Gamma$ is the seeing FWHM. A `---' indicates objects with $r_e 
< 0.6 \, \Gamma$.\\
$^c$Rest-frame colours using SDSS-DR1 PSF magnitudes for the BL Lac
and type 1.x objects, and model magnitudes for all
others. $K$-corrections as above.\\
$^d$A `---' indicates that the H$\alpha$-[\ion{N}{ii}] complex is not
covered.\\
$^e$SF = star-forming galaxy; C = star-forming/AGN composite;
star+gal = superposition of a star and a galaxy.\\
$^f$bc = broad component; A = \citet{Anderson03}; DV =
\citet{devries03}; K = \citet{Kniazev04}; Ka = \citet{Kauffmann03a}; S
= \citet{Schneider05}; Sw = \citet{Schwope00}; W = \citet{Williams02};
RASS = {\em ROSAT} All-Sky Survey \citep{Voges99,Voges00}; FIRST =
\citet{White97}.
\end{minipage}
\end{table}
\end{landscape}

\noindent
We summarise our line ratio measurements and classifications
in columns 7--9 of Table \ref{mcpropstab}. Finally we note that
several of the MCs classified as AGN have been previously classified
by other authors or have been detected in X-ray or radio surveys (see
the last column of Table \ref{mcpropstab}). These findings broadly
support our AGN identifications.

\subsection{Colours}
\label{colprops}
\begin{figure}
\psfig{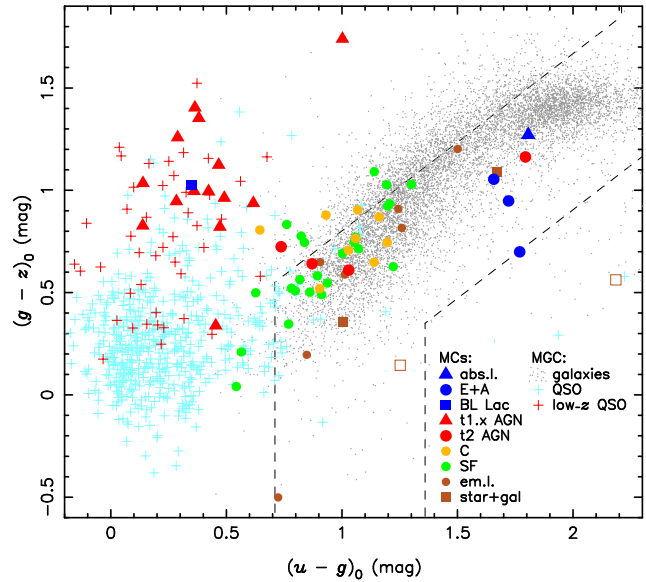}
\caption{Rest-frame colour--colour plot of all $65$ MCs and the general
  MGC galaxy and QSO populations as indicated. Abbreviations are those
  of Table \ref{mcprops}. `abs.l.' and `em.l.' refer to absorption and
  emission line objects without further classification. `low-$z$ QSOs'
  refers to QSOs in the same redshift range as type 1.x AGN, i.e.\
  $0.2 < z < 0.46$. Star/galaxy superpositions are shown with and
  without $K$-corrections (open and solid brown squares respectively).
  We use SDSS-DR1 PSF magnitudes for QSOs, type 1.x AGN and the BL
  Lac, and model magnitudes for all other objects. The dashed lines
  show the (observed-frame) `stellar locus' defined in
  Fig.~\ref{col_sel}.}
\label{col_col}
\end{figure}

In Fig.~\ref{col_col} we compare the MCs to the general MGC galaxy and
QSO populations in the rest-frame $(u-g)_0$--$(g-z)_0$ colour--colour
plane. In this comparison we use SDSS-DR1 PSF magnitudes for QSOs and
for those MCs classified as type 1.x AGN or BL Lac, while we use model
magnitudes for all other objects. For galaxies we also use the
individual $K$-corrections derived by \citet{Driver05}. However, these
will be inappropriate for those MCs whose broad-band fluxes are likely
to be strongly contaminated by an AGN continuum (the type 1.x and BL
Lac objects). For these objects and for QSOs in general we use a crude
$K$-correction assuming a power-law spectral shape with index $\alpha
= -0.5$ \citep{Francis93}. This correction does not account for the
effects of broad emission lines which can contribute substantially to
the flux in a given band. In fact, when considering colours this
procedure is of course equivalent to applying no $K$-correction at
all. The MCs' rest-frame colours are listed in columns 5 and 6 of
Table \ref{mcpropstab}.

The galaxy distribution in Fig.~\ref{col_col} (small grey points)
shows the well-known division between a red ($\approx$ early-type) and
a blue ($\approx$ late-type) population \citep[e.g.][]{Baldry04},
while the QSOs (crosses) exhibit their characteristic UV excess. We
note that low-$z$ QSOs (red crosses) are somewhat redder in $(g-z)_0$
than the bulk of QSOs at higher redshifts (cyan crosses) due to
H$\alpha$ falling in the $z$-band and, presumably, contamination from
the host galaxies \citep[see also][]{Richards01}.

The redshift range of the low-$z$ QSOs was chosen to coincide with
that of the type 1.x AGN (i.e.\ $0.2 < z < 0.46$) and indeed we find
that their colours are matched by the BL Lac (blue square) and by $12$
out of $14$ of our type 1.x AGN (red triangles). This indicates that
the broad-band UV-visual fluxes of these MCs are indeed dominated by
an AGN continuum and not by starlight, supporting our identification
of these objects as AGN based on their emission lines.

One type 1.x (MGC35898) is somewhat bluer in $(g-z)_0$ than the rest,
which is probably due to vigorous star-formation activity which is
indicated by strong, narrow high order Balmer line emission (cf.\
Fig.~\ref{mcspec}e). In contrast, the $14$th type 1.x (MGC03855) is
significantly redder in both colours. However, it is both an X-ray and
radio source and so its identification as an AGN is reasonably
secure. This object may well be reddened by dust in the host galaxy
\citep{Richards03}.

In contrast to type 1.x objects, the AGN contribution to the continua
of type 2 objects is extremely small \citep{Kauffmann03a}, even in
their nuclear regions \citep*{Schmitt99}, and hence their broad-band
colours are expected to be dominated by their host galaxies. Indeed,
the colours of our four type 2 objects (red dots) are quite different
from those of the QSOs but are consistent with normal galaxies.

In a broad-brush sense the properties of the host galaxies of type 2
AGN are known to depend on the AGN's [\ion{O}{iii}] $\lambda 5007$
luminosity \citep{Kauffmann03a}. For example, the mean stellar ages (as
measured by $D_{4000}$ and the H$\delta$ EW) of the hosts of
low-luminosity AGN are similar to those of normal early-type galaxies,
while high-luminosity AGN reside in hosts with younger stellar
populations, similar to those of normal late-type galaxies. This would
explain why we observe both red and blue type 2 objects. To check
whether our objects are consistent with the above trend we need to
measure their [\ion{O}{iii}] luminosities. Unfortunately, only the red
object's spectrum (MGC32111) is flux calibrated: it turns out to be of
intermediate luminosity ($L_{\mbox{\tiny [\ion{O}{iii}]}} =
10^{7.74}L_\odot$) which is however low enough for the $D_{4000}$ and
H$\delta$ EW distributions of type 2s (cf. fig.\ 12 of
\citeauthor{Kauffmann03a}) to still overlap comfortably with the ranges
typical for normal early-type galaxies ($D_{4000} > 1.7$ and H$\delta$
EW $< 1$~\AA). The strength of our object's $4000$~\AA\ break
($D_{4000} = 1.7$) does indeed fall in this overlap region and
indicates the presence of an old stellar population, but its H$\delta$
EW ($= 1.9$~\AA) is somewhat too high for normal early-types. In fact,
MGC32111 is displaced from \citeauthor{Kauffmann03b}'s
(\citeyear{Kauffmann03b}) well-defined $D_{4000}$--H$\delta$
relation. This occurs frequently for type 2s and may be interpreted as
a sign of a significant burst of star-formation in the past two Gyr
\citep{Kauffmann03a}.

An old, evolved stellar population together with a significant,
truncated two-Gyr-old starburst has also been invoked by
\citet{Balogh05} to explain the $(r-K)$ and $(u-g)$ colours of E+A
galaxies, and so it is perhaps not surprising that MGC32111 and our
three E+As (blue dots) are found in a similar region of
Fig.~\ref{col_col}. In this scenario a galaxy in the red locus in
Fig.~\ref{col_col} is moved down and to the left onto the blue
sequence by a burst of star-formation. After the burst's truncation
the galaxy's UV-blue colour reddens again quite rapidly while redder
colours evolve much more slowly, leaving the galaxy below the red
locus in Fig.~\ref{col_col}. We note that our E+As' $(u-g)$ colours
are indeed identical to those found by \citet{Balogh05}.

The only MC that is a normal absorption line galaxy (blue triangle,
cf.\ Fig.~\ref{mcspec}a) lies among the general red galaxy population
as expected.

The MCs identified as star-forming (green dots) or composite (orange
dots) in the previous Section occupy the blue half of the general blue
galaxy sequence. The star-forming MCs extend to its very blue tip. The
three objects furthest along the sequence all show higher order Balmer
lines (e.g.\ MGC35989, see Fig.~\ref{mcspec}k), indicative of strong
star-formation activity.

Notice the odd locations in Fig.~\ref{col_col} of the two likely
star/galaxy superpositions (open brown squares). These are not too
surprising: if the broad-band fluxes of these objects are dominated by
the stars rather than the galaxies then their $K$-corrections are
meaningless. Indeed, the objects' observed-frame colours (solid
brown squares) place them squarely in the centre of the stellar locus
(dashed lines, cf.\ also Fig.~\ref{col_sel}).

The unclassified emission line objects (brown dots) mostly lie along
the general blue sequence. The reddest three objects are also those
with the weakest emission lines while the object near the tip of the
blue sequence (MGC34325) again shows high order Balmer lines. However,
we have re-plotted Fig.~\ref{col_col} using PSF magnitudes and QSO
$K$-corrections for the unclassified objects to see whether any would
lie among the low-$z$ QSOs. This was indeed the case for MGC34325 and
so this object could also be a type 1.x AGN. Finally, the object at
$(g-z)_0 = -0.5$ is the one with saturated H$\alpha$ and
[\ion{O}{iii}] $\lambda5007$. It also shows very strong high order
Balmer lines and may hence be undergoing an extreme starbursting
phase.

\subsection{Redshifts, luminosities and sizes}
\label{zMRprops}
\begin{figure}
\psfig{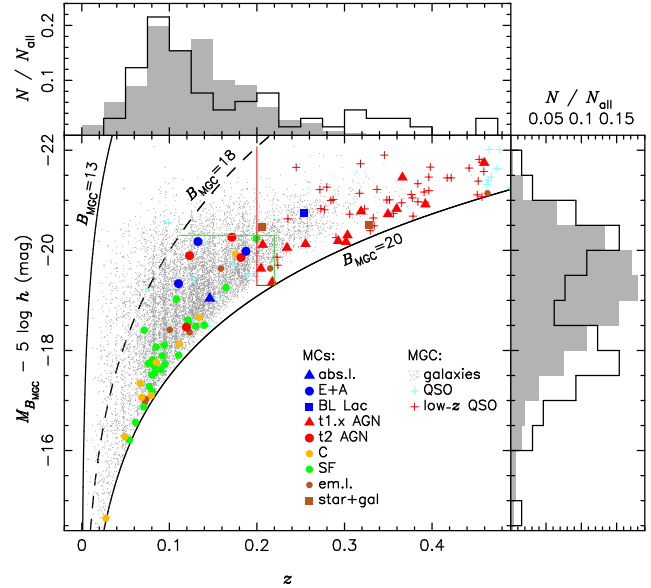}
\caption{Absolute magnitude vs.\ redshift for all $65$ MCs and the
  general MGC galaxy and QSO populations. Symbols are as in
  Fig.~\ref{col_col}. The solid lines show the bright and faint
  apparent magnitude limits of the MGC (using the maximum and minimum
  $K$-correction respectively) while the dashed line marks the
  approximate bright magnitude limit of MCs ($= 18$~mag, cf.\ Section
  \ref{morestel}). The green and red lines show the limits of the
  regions occupied by the `real' and `AGN-dominated' MCs
  respectively. The side panels show the projected normalised
  distributions for the MCs (solid lines) and for the general MGC
  galaxy population (grey shaded histograms).}
\label{Mz}
\end{figure}

In Fig.~\ref{Mz} we show the distribution of the MCs in the absolute
magnitude--redshift plane, along with the rest of the MGC galaxy
population. To calculate the absolute magnitudes we use the same
$K$-corrections as in the previous Section and we list the MCs'
redshifts and magnitudes in columns 2 and 3 of Table \ref{mcpropstab}.

The MCs appear to be segregated in the $M_B$-$z$ plane according to
their type: star-forming and composite objects mostly lie along the
faint magnitude limit at low redshifts, while the type 1.x AGN all lie
at $z > 0.2$ and correspondingly brighter magnitudes. In fact, the
only other MCs at $z > 0.2$ are the BL Lac, the star/galaxy
superpositions and two unidentified emission line objects. One of
these lies at $z \approx 0.2$ and the other is MGC34325 (at $z =
0.46$) which was already suspected of being a type 1.x in the previous
Section. Assuming that MGC34325 is indeed a type 1.x, we find that
{\em all} MCs whose continua are {\em not} dominated by starlight lie
at $z > 0.2$ and $M_B < -19.3$~mag, while the `real' galaxies among
the MCs all lie at $z < 0.22$ and $M_B > -20.3$~mag, so that there is
almost no overlap between the two groups. In particular, this produces
a pronounced bimodality in the MCs' magnitude distribution. From now
on we will generically refer to these two groups as `real' and
`AGN-dominated' MCs ($47$ and $18$ objects respectively, where the
latter group also contains the star/galaxy superpositions).

So why are `real' and `AGN-dominated' MCs distributed so differently?
And why are both groups distributed differently from the general MGC
galaxy population?

For the `AGN-dominated' MCs two obvious explanations come to mind. (i)
It is well-known that the frequency of nuclear activity decreases with
decreasing redshift \citep[e.g.][]{Croom04}. This is also seen in the
MGC where the density of QSOs falls off rapidly towards lower
$z$. (ii) The frequency of nuclear activity might also decrease with
luminosity or size. In this case the hosts of the low-$z$ counterparts
of our type 1.x would probably be resolved and so they would not be
selected as MCs. To test this explanation we would have to fit the
emission lines of all resolved MGC galaxies. However, from
Fig.~\ref{col_col} it is already clear that only very few resolved
galaxies with QSO-like colours exist.

\begin{figure}
\psfig{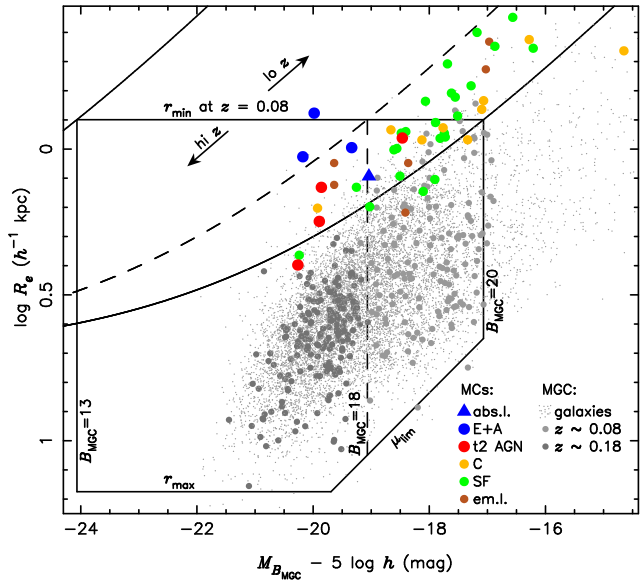}
\caption{Seeing-corrected physical size vs.\ absolute magnitude for
  `real' MCs and the general MGC galaxy population as indicated. The
  large, light and dark grey dots refer only to galaxies in the range
  $18 < \bmgc < 20$~mag. The polygon shows the various selection
  limits at $z = 0.08$: the MGC's apparent magnitude limits, the
  minimum and maximum size limits as well as the low SB limit. The
  vertical dashed line shows the bright limit of the MCs as in
  Fig.~\ref{Mz}. The diagonal lines show how the top part of the
  polygon moves as a function of $z$. The exact locations of the
  polygon and the lines depend on the assumed $K$-correction and
  seeing. Here we have used median values for both.}
\label{RM}
\end{figure}

To understand the distribution of the `real' MCs it is helpful to
consider their magnitude and size selection limits simultaneously. In
Fig.~\ref{RM} we show the `real' MCs along with the rest of the MGC in
the physical size--absolute magnitude plane. To be able to compare the
sizes with each other, especially in the case of small galaxies, we
have applied the seeing correction of \citet{Driver05}. The MCs' sizes
are listed in column 4 of Table \ref{mcpropstab}.

First of all we notice the luminosity--size relation of galaxies:
fainter galaxies tend to be smaller than brighter ones. This already
explains why in Fig.~\ref{Mz} the `real' MCs are mostly found near the
faint magnitude limit: at a given redshift, the smallest galaxies are
also the faintest. However, this should be true at all redshifts and
so this relation by itself does not explain why `real' MCs are
preferentially found at lower $z$.

\begin{figure*}
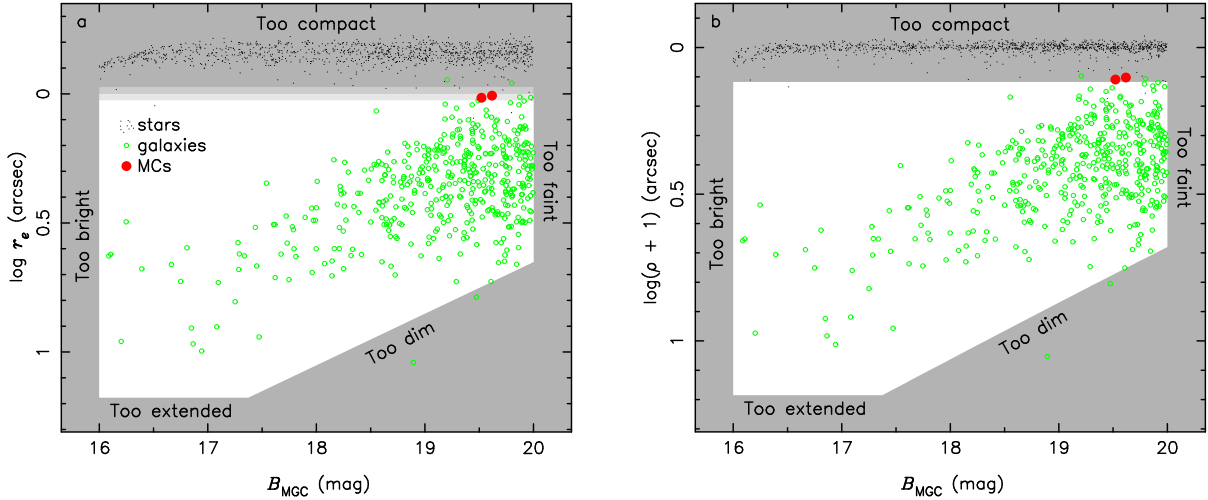

\begin{minipage}{0.9\columnwidth}
\psfig{file=mgc_cg_fig11a.ps,angle=-90,width=\textwidth}
\end{minipage}
\hspace{7mm}
\begin{minipage}{0.9\columnwidth}
\psfig{file=mgc_cg_fig11b.ps,angle=-90,width=\textwidth}
\end{minipage}
\caption{(a) Distribution of objects from the all-object region in the
  observed half-light radius--magnitude plane. Stars and QSOs are
  marked by small black points, galaxies by green open circles and
  misclassified compact galaxies by large solid red dots. Grey shaded
  areas show the observational selection limits discussed in the
  text. The seeing dependence of the minimum size limit is indicated
  by different levels of grey shading. We note in passing that the low
  SB and maximum size limits are {\em not} the absolute detection
  limits but rather the limits to which accurate photometry is
  possible. The slight increase of the stars' $r_e$ at the brightest
  magnitudes is due to the onset of flooding. (b) Same as (a) but
  replacing $r_e$ with $\rho = r_e - 0.6 \, \Gamma$, i.e.\ the
  difference between an object's $r_e$ and the median $r_e$ of stars
  observed in the same seeing.}
\label{r_b}
\end{figure*}

The polygon in Fig.~\ref{RM} encloses the observable parameter space
for galaxies at $z = 0.08$. It is determined by the MGC's apparent
magnitude, size and low SB limits. These selection limits will be
discussed further in Section \ref{mgc_inc} (see Fig.~\ref{r_b} and
accompanying text). The minimum size limit, $r_{\rm min}$, is of
particular interest here because MCs are expected to have apparent
sizes $r_e \la r_{\rm min}$. The three curved diagonal lines in
Fig.~\ref{RM} show how the top part of the polygon moves as a function
of redshift. Essentially, they show the locations of galaxies with
$r_e = r_{\rm min}$ and $\bmgc = 13$, $18$ and $20$~mag respectively
(top left to bottom right). We shall refer to the last of these as the
`small faint line' (SFL). Evidently, galaxies below the SFL cannot
possibly have sizes smaller than $r_{\rm min}$, no matter what their
redshifts are.

The point of this plot is that the slope of the SFL is different from
the slope of the luminosity--size relation of the galaxies. The SFL
actually curves away from the galaxy locus at brighter magnitudes and
hence does not equally cut into the galaxy population at all
luminosities. In fact, at the highest luminosities there is an almost
empty region between the galaxy locus and the SFL and we must conclude
that galaxies are very rare in this part of parameter space. To make
this point clearer we highlight in Fig.~\ref{RM} galaxies at $z
\approx 0.08$ and $z \approx 0.18$ by large light and dark grey points
respectively. For clarity we only show objects with $18 < \bmgc <
20$~mag. The low-$z$ population reaches all the way to $r_{\rm min}$
and so it is not surprising to find galaxies smaller than $r_{\rm
min}$ (and hence `real' MCs) at $z \approx 0.08$. In contrast, the
high-$z$ population does not quite reach to the SFL and hence cannot
reach to $r_{\rm min}$, implying that galaxies do not exist beyond
this limit either (unless the size distribution is bimodal).

Put simply, the distribution of the MCs in Figs.~\ref{Mz} and \ref{RM}
is explained by the fact that the MGC's minimum size limit only
affects faint galaxies and that the MGC is already complete at high
luminosities.

\subsection{Implications}
\label{impl}
In the previous Sections we have seen that the spectra, colours,
magnitudes and redshifts of $18$ of our $65$ MCs indicate that their
broad-band fluxes are not dominated by starlight. In the context of
deriving a luminosity function these objects must be considered a
contamination. In addition, having found them among the MCs raises the
possibility that the non-misclassified compact galaxies are similarly
contaminated. If erroneously included in the analysis these
`AGN-dominated' galaxies could significantly bias our LF estimates.

We have therefore subjected all non-MC galaxies with $r_e < r_{\rm
min} + 0.1$~arcsec
to an analysis of their emission lines and colours along the lines of
Sections \ref{specprops} and \ref{colprops}. There are $192$ such
galaxies but we have found only two type 1.x objects among them. Hence
the high fraction of `AGN-dominated' objects among the MCs must be due
to the colour selection of morphologically stellar targets. In any
case, from now on we will exclude all $20$ `AGN-dominated' compact
galaxies from all further analysis.

\section{MGC incompleteness at high SB}
\label{mgc_inc}
In order to quantify the MGC's incompleteness due to the
misclassification of compact galaxies as stars we will now restrict
ourselves to the complete dataset from the $1.14$~deg$^2$ all-object
region described in Section \ref{aoregion}. $446$ of the $1552$
objects in this region are confirmed galaxies (cf.\ Table
\ref{aotab}). Of these, two objects are MCs, i.e.\ they had originally
been misclassified as stars. (The third MC in this region is
`AGN-dominated' and will not be considered here.)

The goal of this Section is now to use the MCs in the all-object
region to answer the following two questions: (i) How reliable is the
MGC's high SB selection limit (as determined by \citealp{Driver05}),
not just in the all-object region but for the survey as a whole? (ii)
How many galaxies beyond this limit were missed by the MGC?  In other
words we wish to determine the MGC's incompleteness due to
misclassification as a function of SB (or size, see below). Since the
number of misclassified galaxies must depend to some extent on seeing,
this task is complicated, at least in principle, by the fact that the
typical seeing of the all-object region is different from that of the
MGC as a whole. Although the difference is only $10$ per cent we have
no way of knowing a priori whether it is important or not.

We begin by recalling that the MGC's high SB limit is in fact a
minimum size limit \citep[see Appendix A of][]{Driver05}. We use the
half-light radius, $r_e$ (in arcsec), as our size measure, which is
defined as the semi-major axis of the ellipse containing half of the
object's total flux (as measured by {\sc SExtractor}'s BEST
magnitude). In Fig.~\ref{r_b}(a) we show the distribution of objects
from the all-object region in the $r_e$--$\bmgc$ plane. The grey shaded
areas show the various selection limits: the imposed bright and faint
magnitude limits of $16$ and $20$~mag respectively (see Section
\ref{aoregion}), the limiting SB and the maximum and minimum size
limits. We determined the latter three limits by simulating galaxies
over the whole $r_e$--$\bmgc$ plane using the image characteristics of
the real MGC data, and extracting the objects in the same manner as
the real data \citep[see Appendix A of][]{Driver05}. Here we are only
concerned with the minimum size limit, $r_{\rm min}$.

\citet{Driver05} defined $r_{\rm min}$ as the minimum $r_e$ at which a
galaxy can still be {\em reliably} distinguished from stars. To derive
$r_{\rm min}$ they reasoned as follows: the ability to distinguish a
galaxy from stars depends primarily not on the absolute value of the
galaxy's $r_e$ but rather on the distance of the galaxy to the stellar
locus in $r_e$-space, where the scale of that distance is set by the
width of the stellar $r_e$ distribution. Obviously, the observed $r_e$
values of stars and small galaxies are affected by seeing, and hence
$r_{\rm min}$ must be a function of seeing. \citeauthor{Driver05}
observed that for the MGC the median $r_e$ value of stars is given by
$0.6 \, \Gamma$, where $\Gamma$ is the seeing FWHM, and that the width
of the stellar $r_e$ distribution is independent of seeing. These
simple considerations imply $r_{\rm min} = 0.6 \, \Gamma + \Delta$,
and using the simulations they determined $\Delta = 0.31$~arcsec. The
seeing dependence of $r_{\rm min}$ is represented in Fig.~\ref{r_b}(a)
by using different levels of grey shading for the best, median and
worst seeing encountered in the all-object region.

The above also implies that $\rho = r_e - 0.6 \, \Gamma$ is a useful
measure of an object's similarity to stars. In Fig.~\ref{r_b}(b) we
plot the distribution of galaxies and stars in the $\rho$-$\bmgc$
plane. The transformation $r_e \rightarrow \rho$ removes the seeing
dependence of the stellar locus [cf.\ panel (a)], aligning all stars
around $\rho = 0$ while preserving the width of the stellar $r_e$
distribution. The minimum size limit is now simply given by $\rho_{\rm
min} = 0.31$~arcsec and is also seeing independent. Note however that
for a galaxy $\rho$ is {\em not} seeing independent. While a galaxy
moves downwards in Fig.~\ref{r_b}(a) when observed under worse seeing
conditions, it moves upwards in panel (b).

From Fig.~\ref{r_b}(b) we can see that the two MCs lie near $\rho_{\rm
min}$ and other galaxies, which were correctly identified as such,
exist at similar $\rho$ values, indicating that the probability of
misclassifying such galaxies is not $1$. Both MCs lie beyond the
minimum size limit, resulting in $100$ per cent completeness within
the selection limits, while the completeness at $\rho < \rho_{\rm
min}$ is $\sim$$50$ per cent (two out of the four galaxies in this
range were misclassified).

We would now like to estimate the probability of misclassifying a
galaxy and the resulting incompleteness for the full MGC more
quantitatively. First, we define some notation. In the following
$N_{\rm g}$ and $N\dsc{mc}$ denote the numbers of galaxies that were
actually identified as such and of those that were misclassified as
stars respectively. Hence the true number of galaxies is given by
$N\dsc{g} = N\dsc{mc} + N_{\rm g}$. The superscripts `AO' and `MGC'
refer to objects in the all-object region and in the full survey
region respectively.

Now consider the binomial probability, $p_{\rm g}$, that a galaxy is
morphologically misclassified as a star from the imaging data by the
procedure described in Section \ref{mgcdata}. It follows from the
discussion of the minimum size limit above that $p_{\rm g}$ should
only be a function of $\rho$ and that $p_{\rm g}(\rho)$ is the same
for any observation. The number of galaxies that are actually
misclassified in a given observation is then determined by the
distribution of galaxies in $\rho$, $n\dsc{G}(\rho)$, which in turn
depends on the seeing. We would like to estimate the number of
misclassified galaxies in the MGC with $\rho$ larger than some
$\rho_{\rm lim}$:
\begin{equation}
\label{nmc}
N\usc{mgc}\dsc{mc}(\rho_{\rm lim}) = \int_{\rho_{\rm lim}}^\infty
n\usc{mgc}\dsc{mc}(\rho) \, \d\rho = \int_{\rho_{\rm lim}}^\infty p_{\rm
g}(\rho) \; n\usc{mgc}\dsc{g}(\rho) \, \d\rho,
\end{equation}
where we need to measure $p_{\rm g}(\rho)$ from the all-object
region. Ideally, we would estimate it as
\begin{equation}
\label{pg}
p_{\rm g}(\rho) = \frac{n\usc{ao}\dsc{mc}(\rho)}{n\usc{ao}\dsc{g}(\rho)}
\end{equation}
in small bins of $\rho$. However, from Fig.~\ref{r_b}(b) it is clear
that the interesting region of $\rho$ near $\rho_{\rm min}$ is not
populated densely enough to do this. We adopt two approaches to deal
with this.

\subsection{Parameterising $p_{\rm g}(\rho)$}
\label{param_p}
In the following we will assume a functional form for $p_{\rm
g}(\rho)$ and use the data to constrain the model's parameter(s).
Which functional form to use?  Clearly, $p_{\rm g}(\rho)$ should rise
monotonically from $0$ at $\rho \gg \rho_{\rm min}$ to $1$ at $\rho
\ga 0$. Furthermore, given the paucity of data, the model should have
at most two parameters. In general, these two parameters will
characterise in some sense the `position' and steepness of $p_{\rm
g}(\rho)$. Constructing a one-parameter model will require fixing one
of these at some a priori value. The only choice that is not
completely arbitrary seems to be  $p_{\rm g}(0) = 1$.

Using these general considerations as a guideline we now choose the
following four models:
\begin{equation}
\label{plin}
p_{\rm g}^{\rm lin}(\rho) = \left\lbrace
\begin{array}{lcl}
1 & & \qquad \rho \le 0\\
1 - c \rho & \mbox{for} & 0 < \rho < \frac{1}{c}\\
0 & & \frac{1}{c} \le \rho\\ 
\end{array} \right.
\end{equation}
\begin{equation}
\label{pgaus}
p_{\rm g}^{\rm gaus}(\rho) = \left\lbrace
\begin{array}{lcl}
1 & \mbox{for} & \rho \le 0\\
\exp\left(-\frac{\rho^2}{\sigma^2}\right) & & \rho > 0\\
\end{array} \right.
\end{equation}
\begin{equation}
\label{pcos}
p_{\rm g}^{\rm cos}(\rho) = \left\lbrace
\begin{array}{lcl}
1 & & \qquad \rho \le 0\\ 
0.5 \left[1 + \cos\left(\omega \rho \right)\right] & \mbox{for} & 
0 < \rho < \frac{\pi}{\omega}\\
0 & & \frac{\pi}{\omega} \le \rho\\
\end{array} \right.
\end{equation}
\begin{equation}
\label{perf}
p_{\rm g}^{\rm erf}(\rho) = 0.5 \left[1 - {\rm erf}\left(\frac{\rho -
\rho_{50}}{\beta}\right)\right].
\end{equation}
We have chosen three different one-parameter models in order to be
able to test the sensitivity of the results to the precise functional
form. In each case the free parameter ($c$, $\sigma$, $\omega$)
controls the steepness with which $p_{\rm g}$ declines to $0$.  In
contrast we have chosen only one two-parameter model (the error
function) because, as we shall see below, such models are already
over-fitting the sparse data.

\begin{figure}
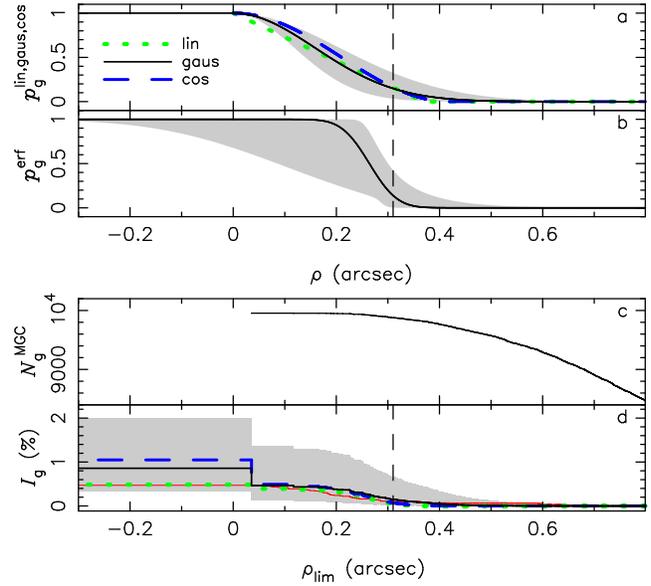

\psfig{file=mgc_cg_fig12a.ps,width=\columnwidth}
\psfig{file=mgc_cg_fig12b.ps,width=\columnwidth}
\caption{(a) Best fit models of the binomial probability of
  misclassifying a galaxy as a function of $\rho = r_e - 0.6 \,
  \Gamma$, using the linear, gaussian and cosine models of equations
  \eref{plin}--\eref{pcos} as indicated. The fits are obtained from
  the observed abundance of MCs in the all-object region, where we can
  be certain that {\em all} galaxies have been identified. The grey
  shaded area shows the $90$ per cent confidence range for the
  gaussian model. (b) Same as (a) for the erf model of equation
  \eref{perf}. (c) Number of observed galaxies (i.e.\ those that were
  originally identified as such) in the full MGC survey region with
  $\rho > \rho_{\rm lim}$. (d) Estimates of the incompleteness of
  galaxies with $\rho > \rho_{\rm lim}$ due to their misclassification
  as stars based on the binomial probabilities in panel (a) and the
  number of galaxies in panel (c). The thin red line is the
  incompleteness derived from all known (`real') MCs in the full MGC
  survey region. Since this sample is incomplete (cf.\ Section
  \ref{morestel}), this line is a strict lower limit to the true
  incompleteness. The grey shaded area shows the $90$ per cent
  confidence range for the gaussian model and the vertical dashed line
  marks the minimum size limit for the reliable detection of galaxies
  adopted by \citet{Driver05}.}
\label{inc_rholim}
\end{figure}

We now fit these models to the data from the all-object region using
maximum likelihood. The likelihood function is given by
\begin{equation}
{\cal L} \propto \prod_{i = 1}^{N\usc{ao}\dsc{mc}} p_{\rm g}(\rho_i)
\; \prod_{j = 1}^{N\usc{ao}_{\rm g}} \, [1 - p_{\rm g}(\rho_j)],
\end{equation}
where the first product runs over all MCs and the second product
over all other galaxies in the all-object region. Maximising this
function with respect to the model parameters yields the results shown
in panels (a)--(b) of Fig.~\ref{inc_rholim}, where the grey-shaded
regions indicate $90$ per cent confidence ranges.

Due to its additional free parameter the erf model behaves quite
differently from the other models. It rises quite steeply just below
$\rho_{\rm min}$ to accommodate the observed misclassified fraction of
$\sim$$50$ per cent in this region (cf.\ Fig.~\ref{r_b}b) and predicts
that essentially all galaxies with $\rho \la 0.18$~arcsec should be
misclassified. The other three models exhibit a more gentle rise, with
the cosine and gaussian models resulting in very similar $p_{\rm g}$
curves. The smallness of the errors on $p_{\rm g}$ for small and large
$\rho$ values is of course a consequence of having assumed a specific
functional form.

We now estimate the number of misclassified galaxies in the full MGC as
\begin{eqnarray}
\label{nmiss}
n\usc{mgc}\dsc{mc}(\rho) & = & p_{\rm g}(\rho) \; n\usc{mgc}\dsc{g}(\rho)
    \nonumber \\ 
    & = & p_{\rm g}(\rho) \left[n\usc{mgc}\dsc{mc}(\rho) +
    n\usc{mgc}_{\rm g}(\rho)\right] \nonumber \\
& = & \frac{p_{\rm g}(\rho)} {1 - p_{\rm g}(\rho)} \; n\usc{mgc}_{\rm
    g}(\rho)
\end{eqnarray}
and define the incompleteness of galaxies with $\rho > \rho_{\rm
lim}$ due to misclassification as
\begin{equation}
\label{inc}
I_{\rm g}(\rho_{\rm lim}) = \frac{N\usc{mgc}\dsc{mc}(\rho_{\rm
lim})}{N\usc{mgc}\dsc{mc}(\rho_{\rm lim}) + N\usc{mgc}_{\rm g}(\rho_{\rm
lim})}.
\end{equation}
$N\usc{mgc}_{\rm g}(\rho_{\rm lim})$ is shown in panel (c) of
Fig.~\ref{inc_rholim}. Recall that according to the erf model
essentially all galaxies with $\rho \la 0.18$~arcsec should have been
misclassified as stars. However, from Fig.~\ref{inc_rholim}(c) we can
see that a few galaxies with $\rho < 0.18$~arcsec were in fact
identified in the full survey. This causes the erf model to predict an
infinite number of misclassified galaxies. This unrealistic result is
of course due to the paucity of data in the all-object region compared
to the number of free parameters of the erf model, which is also
reflected by the large error on $p_{\rm g}^{\rm erf}$ shown in in
Fig.~\ref{inc_rholim}(b). We will not consider this model any further.

Note also that the few observed galaxies at small $\rho$ values also
justify our choice of $p_{\rm g}(0) = 1$ for the one-parameter
models. Moving the $p_{\rm g} = 1$ point to a larger $\rho$ value
would have caused the same problem as above.

The incompleteness resulting from the linear, gaussian and cosine
models is shown in panel (d) of Fig.~\ref{inc_rholim}. For a given
model we can now read off the numbers that were the goal of this
Section: (i) the incompleteness of galaxies within the selection
limit, i.e.\ at $\rho_{\rm lim} = \rho_{\rm min}$ where the MGC was
assumed to be complete by \citet{Driver05} and (ii) the MGC's overall
incompleteness. We summarise these values in columns 2 and 4 of Table
\ref{inctab} respectively. We also show in column 6 the incompleteness
of galaxies beyond the selection limit.

From Fig.~\ref{inc_rholim}(d) we can see that the differences between
the three models are smaller than the random errors (shown only for
the gaussian model by the grey shaded area). Hence, given the size of
the errors, the results do not depend sensitively on the precise shape
of the assumed functional form of $p_{\rm g}$.

Nevertheless, the incompleteness estimate provided by the linear model
is somewhat lower than that of the other two models and may in fact be
too low. In Fig.~\ref{inc_rholim}(d) we plot as a thin red line the
incompleteness derived from {\em all known} `real' MCs in the MGC
($47$ in total), i.e.\ those in the all-object region as well as those
discovered through our additional spectroscopy of stellar objects
described in Section \ref{morestel}. Since this sample is itself
incomplete, the incompleteness derived from it must be considered a
strict lower limit to the true incompleteness. The estimate from the
linear model traces this limit almost exactly.
The cosine model, which gives the highest incompleteness estimates, is
most consistent with the lower boundary. We summarise the lower limit
on the incompleteness in the last line of Table \ref{inctab}.

\begin{table*}
\begin{minipage}{14cm}
\caption{Estimates of the MGC's incompleteness due to misclassification of
galaxies as stars ($16 \le \bmgc < 20$~mag).}
\label{inctab}
\centerline{\begin{tabular}{lrrcrrcrr}
\hline
\hspace{2cm} & \multicolumn{2}{c}{$\rho_{\rm min} < \rho$} & \hspace{5mm} &
\multicolumn{2}{c}{$0 < \rho$} & \hspace{5mm} &
\multicolumn{2}{c}{$0 < \rho < \rho_{\rm min}$}\\
& $I_{\rm g}(<)^a$ & $R_{\rm g}^b(<)^a$ & & $I_{\rm g}(<)^a$ &
$R_{\rm g}^b(<)^a$ & & $I_{\rm g}(<)^a$ & $R_{\rm g}^b(<)^a$\\
& \multicolumn{2}{c}{(per cent)} & & \multicolumn{2}{c}{(per cent)} & &
\multicolumn{2}{c}{(per cent)}\\
\hline 
\multicolumn{4}{l}{From Section \ref{param_p}, using different models:}\\
Linear   & $0.05(0.51)$ & $    (0.41)$ & & $0.49(1.28)$ & $0.02(0.81)$
& & $36.0(50.5)$ & $ 5.2(26.7)$\\
Gaussian & $0.15(0.66)$ & $0.05(0.56)$ & & $0.86(2.01)$ & $0.39(1.54)$
& & $48.1(64.2)$ & $23.2(47.0)$\\
Cosine   & $0.07(0.51)$ & $    (0.41)$ & & $1.05(2.31)$ & $0.58(1.85)$
& & $56.1(70.5)$ & $35.0(56.4)$\\
\\
\multicolumn{4}{l}{From Section \ref{noseediff}, based on different objects:}\\
galaxies        & $0(0.52)$ & $(0.42)$ & & $0.45(1.40)$ & $    (0.95)$ & &
$35.8(52.8)$ & $6.6(31.3)$\\
stellar objects & $0(0.93)$ & $(0.83)$ & & $0.65(1.98)$ & $0.19(1.52)$ & & 
$44.9(57.4)$ & $19.8(38.0)$\\
visually inspected & $0(0.99)$ & $(0.90)$ & & $1.07(2.94)$ & $0.61(2.49)$ & & 
$57.1(71.5)$ & $37.6(58.6)$\\
Area & $0(0.82)$ & $(0.72)$ & & $0.54(1.69)$ & $0.07(1.23)$ & 
& $41.4(53.8)$ & $13.2(31.6)$\\
\\
From Section \ref{incmc} & \multicolumn{1}{l}{$0.14$} &
\multicolumn{1}{l}{$0.04$} & & \multicolumn{1}{l}{$0.61$} & 
\multicolumn{1}{l}{$0.15$} & & \multicolumn{1}{l}{$38.6$} &
\multicolumn{1}{l}{$9.0$}\\
Lower limit$^c$ & \multicolumn{1}{l}{$0.10$} & & & 
\multicolumn{1}{l}{$0.47$} & & & \multicolumn{1}{l}{$32.5$}\\
\hline
\end{tabular}}
$^a$Upper limit at $95$ per cent confidence.\\
$^b$Remaining incompleteness after accounting for `real' MCs already known.\\
$^c$Derived from all known `real' MCs in the MGC.
\end{minipage}
\end{table*}

Note the tail of MCs at large $\rho$ values identified in the
additional data. There are a total of $10$ MCs with $\rho > \rho_{\rm
min}$, however the tail consists mostly of six objects at $\rho >
0.45$~arcsec. Their large $\rho$ values are due to $r_e$ measurement
errors caused by slightly erroneous {\sc SExtractor} object ellipses
which in turn are due to nearby bright objects in four of the six
cases. The one-parameter models are not equipped to handle a non-zero
tail of objects at large $\rho$. The gaussian is best suited, but even
for this model the incompleteness derived from these objects
eventually exceeds even the upper end of the $90$ per cent confidence
range. This highlights that the derived errors on the incompleteness
estimates do not include the shortcomings of the chosen models.

In columns 3, 5 and 7 of Table \ref{inctab} we list the MGC's
remaining incompleteness, $R_{\rm g}(\rho_{\rm lim})$, after
accounting for those MCs that have already been discovered.
Specifically, $R_{\rm g}$ is calculated by applying equation
\eref{inc} after having subtracted the known `real' MCs from the
number of MCs predicted from the all-object region
($N\usc{mgc}\dsc{mc}$) in the numerator. If the number of known MCs
exceeds the predicted number (see discussion above) we only list an
upper limit.

\subsection{Assuming $\Gamma\dsc{ao} = \Gamma\dsc{mgc}$}
\label{noseediff}
Recall that the formalism of the previous Section was necessary
because of the different seeing in the all-object region and the MGC
as a whole. In this Section we will simply ignore this
difference. Since in fact $\Gamma\dsc{AO} < \Gamma\dsc{MGC}$, this
assumption introduces a bias towards lower incompleteness.

Ignoring the seeing difference means that we can assume that the
$\rho$ distribution of galaxies in the all-object region is the same
as in the full MGC (apart from the normalisation):
\begin{equation}
n\usc{ao}\dsc{g}(\rho) = n\usc{mgc}\dsc{g}(\rho)
\frac{N\usc{ao}\dsc{g}}{N\usc{mgc}\dsc{g}},
\end{equation}
where $N = \int_{-\infty}^\infty n(\rho) \, \d\rho$. Substituting into
equations \eref{pg} and \eref{nmc} we find for the number of
misclassified galaxies in the full MGC:
\begin{equation}
\label{Nmiss}
N\usc{mgc}\dsc{mc}(\rho_{\rm lim}) = N\usc{ao}\dsc{mc}(\rho_{\rm lim})
\frac{N\usc{mgc}\dsc{g}}{N\usc{ao}\dsc{g}},
\end{equation}
i.e.\ we simply scale the number of MCs in the all-object region by
the ratio of the total numbers of galaxies in the full MGC and the
all-object region. For simplicity and so that we do not have to make
any reference to objects with $\rho < \rho_{\rm lim}$ we prefer to
replace this ratio with $N\usc{mgc}\dsc{g}(\rho_{\rm lim}) /
N\usc{ao}\dsc{g}(\rho_{\rm lim})$ which is valid under the assumption
of having the same seeing in the all-object region and in the full
MGC. In analogy to equation \eref{pg} we now define the `cumulative'
binomial probability as
\begin{equation}
\tilde p_{\rm g}(\rho_{\rm lim}) = \frac{N\usc{ao}\dsc{mc}(\rho_{\rm
    lim})}{N\usc{ao}\dsc{g}(\rho_{\rm lim})}
\end{equation}
and we rewrite equation \eref{Nmiss} as
\begin{eqnarray}
N\usc{mgc}\dsc{mc}(\rho_{\rm lim}) & = & 
    \tilde p_{\rm g}(\rho_{\rm
    lim}) N\usc{mgc}\dsc{g}(\rho_{\rm lim}) \nonumber\\ 
    & = & \tilde p_{\rm
    g}(\rho_{\rm lim}) \left[N\usc{mgc}\dsc{mc}(\rho_{\rm lim}) +
    N\usc{mgc}_{\rm g}(\rho_{\rm lim})\right] \nonumber\\
& = & \frac{\tilde p_{\rm g}(\rho_{\rm lim})}
{1 - \tilde p_{\rm g}(\rho_{\rm lim})}
N\usc{mgc}_{\rm g}(\rho_{\rm lim}),
\end{eqnarray}
analogous to equation \eref{nmiss}.

Given the assumption $\Gamma\dsc{ao} = \Gamma\dsc{mgc}$ we can
consider several other ways of estimating $N\usc{mgc}\dsc{mc}$:

First, instead of scaling the number of MCs by the total number of
galaxies as in equation \eref{Nmiss}, we can simply scale by the
observed areas on the sky, $A$, thus making no reference to any
objects other than the MCs:
\begin{equation}
N\usc{mgc}\dsc{mc}(\rho_{\rm lim}) = N\usc{ao}\dsc{mc}(\rho_{\rm lim})
\frac{A\usc{mgc}}{A\usc{ao}}
\end{equation}

Next, from an a posteriori point of view it is interesting to ask:
given an object morphologically classified as stellar, what is the
binomial probability, $\tilde p_{\rm s}$, that it is actually a
galaxy? We now have
\begin{equation}
\tilde p_{\rm s}(\rho_{\rm lim}) = \frac{N\usc{ao}\dsc{mc}(\rho_{\rm
    lim})}{N\usc{ao}_{\rm s}(\rho_{\rm lim})}
\end{equation}
and
\begin{equation}
\label{Nmisss}
N\usc{mgc}\dsc{mc}(\rho_{\rm lim}) = \tilde p_{\rm s}(\rho_{\rm lim})
\; N\usc{mgc}_{\rm s}(\rho_{\rm lim}),
\end{equation}
where $N_{\rm s}$ is the number of objects morphologically classified
as stellar. This includes stars, QSOs and MCs. Note that the process
of testing whether a morphologically stellar object is in fact a
galaxy is not really a binomial process because the outcome is
pre-determined. In fact, compared to the previous two estimates of
this Section we expect equation \eref{Nmisss} to result in a higher
incompleteness because the average stellar density in the full survey
is higher than in the all-object region (due to different mean
Galactic latitudes, cf.\ fig.~11 of \citealp{Liske03b}).

Finally, we note that both of the MCs in the all-object region
have stellaricity $< 0.98$ (cf.\ Fig.~\ref{mcimg}) and therefore their
original classification as stars was the result of visual inspection
(see Section \ref{mgcdata}). Hence we can gain yet another estimate of
the incompleteness by substituting $N_{\rm s}$ above with the
respective number of stellar objects classified as such by visual
inspection, $N_{\rm v}$. However, we must expect this estimate to
return an even higher incompleteness than the previous one because it
suffers an additional bias: the fraction of stars classified visually
is higher in the full survey than in the all-object region because of
the worse seeing.


\begin{figure}
\psfig{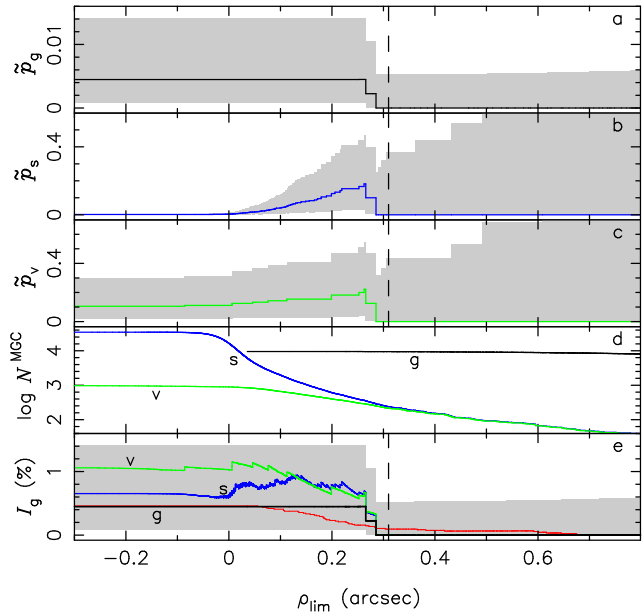}
\caption{(a) Cumulative binomial probability of misclassifying a
  galaxy with $\rho > \rho_{\rm lim}$ as a star. (b) Cumulative
  binomial probability that an object classified as stellar and with
  $\rho > \rho_{\rm lim}$ is actually a galaxy. (c) Cumulative
  binomial probability that an object classified as stellar by visual
  inspection and with $\rho > \rho_{\rm lim}$ is actually a
  galaxy. (d) Number of objects in the full MGC survey region with
  $\rho > \rho_{\rm lim}$: `g' and `s' refer to objects originally
  classified as galaxies and stars, while `v' denotes that subset of
  `s' which was classified by eye. (e) Estimates of the incompleteness
  of galaxies with $\rho > \rho_{\rm lim}$ due to their
  misclassification as stars based on the probabilities in panels
  (a)--(c) (which were derived from the all-object region) and the
  numbers of objects in panel (d) as indicated. The thin red line is
  the incompleteness derived from all known (`real') MCs in the full
  MGC survey region, and gives a lower limit on the true
  incompleteness. The grey shaded areas show $90$ per cent confidence
  ranges and the vertical dashed line marks the minimum size limit
  adopted by \citet{Driver05}.}
\label{inc_rholim2}
\end{figure}

In Fig.~\ref{inc_rholim2} we show the above cumulative probabilities
derived from the all-object region (panels a--c), the appropriate
numbers of objects in the full survey region (d) and the resulting
incompleteness estimates (e). The latter are summarised in Table
\ref{inctab}.

Given the errors, the derived $I_{\rm g}$ values of this Section are
reasonably consistent with each other. This implies that the effects
of the differences in the stellar density and visually classified
fraction between the all-object region and the full MGC are not
dramatic. Nevertheless, they are detectable, and we have $I_{\rm g} <
I_{\rm g}^{\rm s}< I_{\rm g}^{\rm v}$ as expected. We also find that
$I_{\rm g}$ is smaller than the incompleteness estimates from the
previous Section and we attribute this to the seeing difference
between the all-object region and the full MGC. However,
compared to the size of the random errors the systematic shift is
again small.

Note that both MCs in the all-object region have $\rho < \rho_{\rm
min}$. Therefore the all-object region predicts $I_{\rm g}(\rho_{\rm
lim} = \rho_{\rm min}) = 0$ which is of course below the lower limit
set by the sample of already known MCs in the full MGC (again shown as
a thin red line in Fig.~\ref{inc_rholim2}e). Hence, in Table
\ref{inctab} we only list upper limits for the residual incompleteness
$R_{\rm g}$ at $\rho_{\rm min}$.

A major difference between Figs.~\ref{inc_rholim2}(e) and
\ref{inc_rholim}(d) is the size of the errors for large $\rho_{\rm
lim}$. The error of, say, $I_{\rm g}(0.6)$ in
Fig.~\ref{inc_rholim2}(e) is simply set by the observation that none
out of $N\usc{mgc}\dsc{g}(0.6)$ galaxies in the all-object region had
been misclassified, while the smallness of the error on $I_{\rm
g}^{\rm gaus}(0.6)$ in Fig.~\ref{inc_rholim}(d) is due to the
additional assumption of a specific function for $p_{\rm g}$.

Even though $I_{\rm g}(\rho_{\rm lim})$ should be a monotonic
function, $I_{\rm g}^{\rm s}$ in Fig.~\ref{inc_rholim2}(e) exhibits a
small decline from $\rho_{\rm lim} \approx 0.12$ to $0$~arcsec. This
reflects a small misalignment between the stellar $\rho$
distributions, $n_{\rm s}(\rho)$, in the all-object region and the
full survey. In any case, since the fluctuation is much smaller than
the errors, it has little effect on the result.

\subsection{Correcting the MC sample for incompleteness}
\label{incmc}
In the bottom panels of Figs.~\ref{inc_rholim} and \ref{inc_rholim2}
we have compared the various $I_{\rm g}$ estimates predicted from the
{\em complete} sample of MCs in the all-object region with the lower
limit on $I_{\rm g}$ derived from the {\em incomplete} sample of all
known `real' MCs in the full MGC. Almost all of the estimates are
within a factor of $2$ of the lower limit, suggesting that our
additional observations of stellar objects have probed at least part
of that region of observational parameter space in which MCs exist.

As a cross-check we will now apply a rudimentary correction to the
number of MCs discovered outside of the all-object region (see Section
\ref{morestel}) to see whether we recover a similar overall
incompleteness as in the previous Sections. We perform the correction
by dividing the four-dimensional parameter space spanned by $\bmgc$,
$(u-g)$, $(g-z)$ and SDSS-DR1 morphological classification into five
separate regions. We then determine the spectroscopic incompleteness
of stellar objects in each of these regions and apply this factor to
the number of MCs discovered in them, implicitly assuming that each
region was sampled homogeneously. Since MGCz contributed most of the
spectra in the `additional' sample, we follow the MGCz target
selection described in Section \ref{morestel} in defining the above
regions. The result of applying this correction is that the MGC's
overall incompleteness is $0.61$ per cent, in reasonable agreement
with the models of Section \ref{param_p} (cf.\ column 4 of Table
\ref{inctab}). The $95$ per cent confidence upper limit on this number
is $1.74$ per cent but there is an unknown additional uncertainty due
to our use of an approximate selection function. Hence the estimate
above is much less reliable than those of the previous Sections and we
do not quote any upper limits in Table \ref{inctab}.

\section{Luminosity function and density including compact galaxies}
\label{cglf}
In the previous Section we have seen that the standard MGC galaxy
catalogue is missing up to $64$ per cent of galaxies with $r_e <
r_{\rm min}$. However, since this translates to only $2$ per cent of
{\em all} galaxies (cf.\ columns 6 and 4 of Table \ref{inctab},
gaussian model) one might be tempted to conclude that MCs and other
compact galaxies are entirely negligible with respect to the galaxy
population as a whole. However, since the raw number of non-compact
galaxies is dominated by luminous objects and since compact galaxies
are preferentially faint (cf.\ right panel of Fig.~\ref{Mz}), the
above impression may be misleading. Clearly, we need to consider the
issue as a function of luminosity before we can reach a final
conclusion regarding the relevance of compact galaxies and their
misclassification. The goal of this Section is therefore to evaluate
the contribution of compact galaxies to the local galaxy LF.

In the following we construct LFs using a variant of the bivariate
step-wise maximum likelihood (SWML) method of \citet{Driver05}. We
impose the same apparent magnitude, low-SB and redshift limits: $13 <
\bmgc < 20$~mag, $r_e < 15$~arcsec, $\mu_{\rm eff} < 25.25$\mpass\ and
$0.013 < z < 0.18$, but we do {\em not} impose a minimum size
limit. In our context of compact galaxies we can hence regard this
method as being equivalent to a standard monovariate SWML analysis
\citep*{Efstathiou88}, where all SB selection effects are ignored. The
only difference is that we are in fact accounting for the MGC's low-SB
limits.

The application of the above limits leaves us with $7803$
non-misclassified galaxies. To convert apparent to absolute magnitudes
we use the \citeauthor{Driver05} global evolutionary correction and
individual $K$-corrections, where the former is given by $E(z) = -0.75
\times 2.5 \log(1+z)$.

\begin{figure}
\psfig{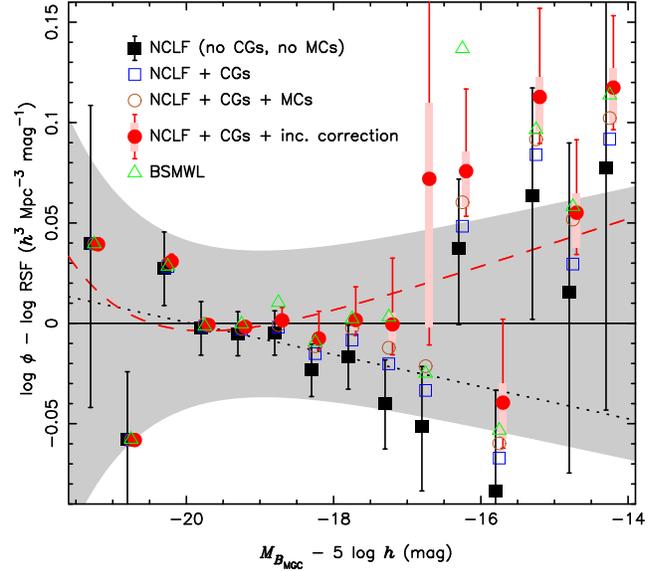}
\caption{Various estimates of the MGC's luminosity function as
  indicated and explained in the text. All estimates are normalised by
  an arbitrary reference Schechter function (RSF), which has $M_B^* -
  5 \log h = -19.63$~mag, $\phi^* = 0.0174$~$h^3$~Mpc$^{-3}$ and
  $\alpha = -1.16$. The grey shaded area shows a $\pm 1\sigma$ range
  around the RSF that is typical of the Schechter function fits of the
  various LFs. The LFs are derived using an essentially standard SWML
  method, except the one labelled `BSWML' where we use the full
  bivariate method of \citet{Driver05}. `CGs' refers to
  non-misclassified compact galaxies. The errors on the NCLF are
  poissonian only. The errors on the incompleteness corrected LF
  reflect the uncertainties due to the choice of the incompleteness
  model (boxes) and the models' parameters (errorbars).  The black
  dotted and red dashed lines show the best fit Schechter function for
  the NCLF and the incompleteness corrected LF respectively. Some data
  points are offset horizontally for clarity.}
\label{alllfs}
\end{figure}

We begin now by first constructing the LF of non-compact galaxies
only, i.e.\ we exclude all MCs as well as all non-misclassified
galaxies with $r_e < r_{\rm min}$ from the sample. This is the sample
that would result from a severe high-SB selection effect (in the sense
that {\em every} galaxy below $r_{\rm min}$ is missed) and its LF will
obviously be an under-estimate of the true LF. In the following we
will refer to this as the non-compact LF (NCLF). In Fig.~\ref{alllfs}
we show the NCLF as black solid squares, where we have normalised all
data points by an arbitrarily chosen \citet{Schechter76} function with
parameters: $M_B^* - 5 \log h = -19.63$~mag, $\phi^* =
0.0174$~$h^3$~Mpc$^{-3}$ and $\alpha = -1.16$.  We show only poisson
errors since we will only be comparing various estimates of the LF
derived from essentially the same volume.

Next we add in the non-misclassified compact galaxies with $r_e <
r_{\rm min}$ (CGs), which somewhat raises the LF at the faint end as
expected (blue open squares). However, we know that these CGs are
$\sim$$50$ per cent incomplete, and so we are still under-estimating
the true LF. This is the LF one obtains from the MGC data if the issue
of high-SB selection effects is simply ignored.

The brown open circles in Fig.~\ref{alllfs} show the result of also
adding in the known `real' MCs. Since not all MCs were actually
discovered, this sample is still incomplete and we are still
under-estimating the true LF.

Finally, we derive a LF estimate using an incompleteness correction:
in Section \ref{param_p} we postulated that the classification of
galaxies is a random process, where the binomial probability of
misclassifying a galaxy as a star, $p_{\rm g}$, depends only on the
galaxy's $\rho = r_e - 0.6 \, \Gamma$. Hence we can recover the total
number of galaxies by simply weighting each non-misclassified galaxy
by $[1 - p_{\rm g}(\rho)]^{-1}$. The result is shown as red solid dots
in Fig.~\ref{alllfs} where we have used the gaussian estimate of
$p_{\rm g}$ (cf.\ Section \ref{param_p}). Using the cosine or linear
estimates instead results in the LFs marked by the upper and lower
ends of the light red vertical boxes respectively. The errorbars
protruding from these boxes show the uncertainties due to the $90$
percentile errors on the cosine and linear model parameters.

We list the parameters of the best fitting Schechter functions of the
above LFs in Table \ref{lftab}.

For comparison we also plot in Fig.~\ref{alllfs} as green open
triangles the MGC LF as derived by using the fully bivariate SWML
method of \citet{Driver05}, including the minimum size limit. In this
case we are using the same sample as for the NCLF, i.e.\ the MCs and
CGs are excluded, but now the $r_{\rm min}$ selection limit is
properly taken into account in the SWML analysis by appropriately
reducing each galaxy's observable parameter space in the
luminosity--SB plane. This is equivalent to reducing the volume over
which a galaxy could have been observed, and hence the resulting LF is
somewhat higher than in the standard NCLF case.

Examining Fig.~\ref{alllfs} we first of all notice that measurable
differences between the various LF estimates are confined to the
regime $M_B \ga -18$~mag. As mentioned above, this is due to the
combined effect of $r_{\rm min}$ and the luminosity--size relation of
galaxies as discussed in connection with Fig.~\ref{RM}.

Secondly, comparing the red dots to the blue open squares, it seems
that the incompleteness of compact galaxies due to misclassification
does not affect the LF significantly: for almost all points the
differences are smaller than the $1\sigma$ poisson error. In this
sense even the effect of the {\em total} compact galaxy population is
only marginally significant (comparing the red dots with the black
squares). However, the impact of the compact galaxies is obviously
systematic and hence more significant when considering the faint end
of the LF as a whole: the overall effect of the compact galaxies is to
steepen the faint-end slope $\alpha$ by $0.05$, which is a $2.5\sigma$
effect (cf.\ Table \ref{lftab}), while the misclassified galaxies
alone steepen $\alpha$ by $0.03$, which is still marginally
significant. The steepening increases to $0.05$ if we use the upper
end of the $90$ percentile confidence range on $p_{\rm g}^{\rm
gaus}(\rho)$, and to $0.06$ if we use the most extreme incompleteness
correction (cf.\ last line of Table \ref{lftab}).

\begin{figure}
\psfig{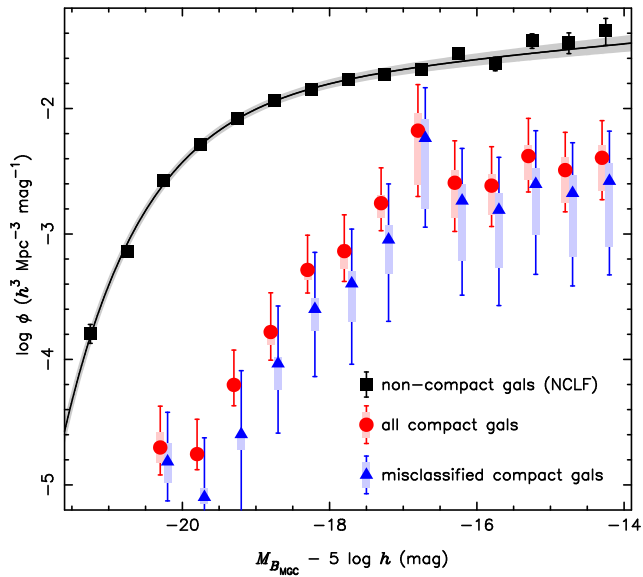}
\caption{LFs of compact and non-compact galaxies as indicated. The
  black line and grey region show the Schechter function that best
  fits the NCLF, and $\pm 1\sigma$ range. All errorbars have the same
  meaning as in Fig.~\ref{alllfs}.}
\label{clfs}
\end{figure}

To further illustrate this point, we directly compare the LFs of
compact and non-compact galaxies in absolute terms in
Fig.~\ref{clfs}. Although there are $\sim$$2$ orders of magnitude
fewer compact galaxies than non-compact ones at bright magnitudes,
they make up $\sim$$10$ per cent of all galaxies in the range $-17 <
M_B < -14$~mag, and the misclassified galaxies alone contribute
$\sim$$6$ per cent in this range.

Finally, comparing the incompleteness corrected LF to the fully
bivariate SWML estimate in Fig.~\ref{alllfs} (green triangles) we find
that the two agree very well for almost all points. Put simply, the
first method corrects for missing objects while the second method
corrects for missing parameter space (or volume), and it is reassuring
that the two methods yield the same result with no obvious systematic
bias.

Since the luminosity density, $j$, is dominated by the bright end of
the LF it is only marginally affected by the compact galaxies. For a
given LF we calculate the luminosity density from the parameters of
the best fit Schechter function as $j = \phi^* \, L^*_{\bmgc} \,
\Gamma(\alpha +2)$. From the last column of Table \ref{lftab} we can
see that the inclusion of compact galaxies increases $j$ by $3.5$ per
cent, while the misclassified galaxies alone are responsible for an
increase of only $2$ per cent.

\begin{table*}
\begin{minipage}{12.9cm}
\caption{Schechter function parameters for the LFs of
Fig.~\ref{alllfs}.}
\label{lftab}
\centerline{\begin{tabular}{lllll}
\hline
Sample & \multicolumn{1}{c}{$\phi^*$} & \multicolumn{1}{c}{$M_B^* - 5
\log h$} & \multicolumn{1}{c}{$\alpha$} & \multicolumn{1}{c}{$j$}\\ &
\multicolumn{1}{c}{($10^{-2} h^3$ Mpc$^{-3}$)} &
\multicolumn{1}{c}{(mag)} & & \multicolumn{1}{c}{($10^8 h L_{\odot}$
Mpc$^{-3}$)}\\
\hline 
NCLF & $1.73 \pm 0.07$ & $-19.63 \pm 0.04$ & $-1.14 \pm 0.02$ & $1.93 \pm 0.13$\\
NCLF+CGs                         & $1.72$ & $-19.64$ & $-1.16$ & $1.97$\\ 
NCLF+CGs+MCs                     & $1.70$ & $-19.64$ & $-1.17$ & $1.96$\\
NCLF+CGs+inc.\ correction        & $1.67$ & $-19.66$ & $-1.19$ & $2.00$\\
NCLF+CGs+extreme inc.\ cor.$^a$  & $1.64$ & $-19.68$ & $-1.22$ & $2.06$\\
\hline
\end{tabular}}
$^a$Using the upper end of the $90$ percentile confidence range of
the cosine estimate of $p_g(\rho)$.\\
\end{minipage}
\end{table*}

\section{Conclusions}
\label{conclusions}
We have explored the very high surface brightness (SB) regime of the
Millennium Galaxy Catalogue (MGC) to assess the impact of compact
galaxies on the field galaxy luminosity function (LF) in the local ($z
\approx 0.1$) Universe. We have observationally defined `compact
galaxies' as objects whose half light-radii are so small that they
cannot {\em reliably} be distinguished from stars in the MGC using
{\sc SExtractor} and/or visual examination (i.e.\ $r_e \la 1$~arcsec).

We have studied the incompleteness of the standard MGC galaxy
catalogue ($\bmgc < 20$~mag) due to the misclassification of compact
galaxies as stars using the $1.14$~deg$^2$ all-object sub-region of
the MGC, where we have spectroscopically identified {\em all} galaxies
in the range $16 < \bmgc < 20$~mag. We find:


1. Within the MGC's minimum size limit (as derived by
   \citealp{Driver05}) only $0.15$ per cent of galaxies are missed
   due to their compactness, with a $95$ per cent upper limit of
   $0.66$ per cent. Hence we verify \citeauthor{Driver05}'s assumption
   that the MGC is complete within its minimum size limit.

2. Beyond the minimum size limit $\sim$$50$ per cent of galaxies are
   misclassified as stars.

3. However, with respect to the total galaxy population this
   translates to an incompleteness of only $1.05$ per cent of all
   galaxies, with a $95$ per cent upper limit of $2.31$ per
   cent.


Our incompleteness is a factor of $2.6$ lower than the $(2.8 \pm 1.6)$
per cent found by \citet{Drinkwater99} for galaxies with $16.5 \le
b_{\rm J} \le 19.7$~mag in the Fornax Cluster Spectroscopic Survey
(FCSS). This is presumably due to the higher resolution of the MGC
compared to the UK Schmidt photographic data from which the FCSS was
selected. The null-result of \citet{Morton85} on the other hand can be
explained by their small survey area of $0.31$~deg$^2$.

The incompleteness values above ignore the fact that a total of $47$
misclassified compact galaxies (MCs) have already been discovered
through additional spectroscopy of morphologically stellar
objects. Accounting for these objects reduces the overall
incompleteness to just $0.58$ per cent.

Using this extended, albeit incomplete sample of $47$ objects we have
studied the nature of the MCs. Over $90$ per cent of them are emission
line galaxies (similar to the fraction of \citealp{Drinkwater99}), but
we also find three E+A galaxies. Although some E+As show clear signs
of interaction \citep{Zabludoff96,Yang04} they are generally found to
be bulge-dominated, spheroidal systems
\citep{Quintero04,Yang04,Blake04} with bright compact cores
\citep{Goto05} and higher central SB than normal bulge-dominated
galaxies \citep{Quintero04,Yang04}. Hence their appearance among the
MCs is not too surprising.

An analysis of emission line ratios revealed that four of the emission
line galaxies harbour a type 2 AGN. Even though early studies of
Seyfert galaxies seemed to suggest that they were mostly hosted by
spiral galaxies \citep{Adams77}, recent results from the SDSS have
shown that the surface mass density distribution of type 2 AGN hosts
is very similar to that of normal early-type galaxies
\citep{Kauffmann03a}. Also, their concentration index distribution
extends to very high values and so we conclude that our compact type
2s are not unusual.

Blue compact galaxies, although ill-defined, are the most common and
widely studied class of compact galaxies (see \citealp{Kunth00} for a
review) and we find that $83$ per cent of our MCs are blue,
star-forming galaxies with strong nebular emission lines. We note that
these galaxies generally have $M_B > -19$~mag and are hence fainter
than the luminous blue compact galaxies
\citep[e.g.][]{Phillips97,Werk04} but overlap with the blue compact
dwarfs \citep[BCDs; e.g.][]{Thuan81,GildePaz03}.

The `blue spheroids' of \citet{Ellis05} \citep[see
also][]{Cross04a,Driver06} are another type of blue, star-forming,
sub-$L^*$ galaxy. However, in contrast to the BCDs, which often have
disturbed or irregular morphologies \citep{Cairos01}, the blue
spheroids are very smooth, with concentration and asymmetry indeces
similar to normal ellipticals. Since our MCs are only barely resolved
we have not been able to reliably fit their SB profiles. Hence,
without higher resolution imaging it is not possible to decide whether
the MCs are more similar to BCDs or a high SB version of the blue
spheroids.



The faintness of the MCs, which is essentially due to the
luminosity--size relation of galaxies, is the cause of their
differential effect on the local galaxy luminosity function: our main
result is that even in the relatively high-resolution MGC the
misclassification of compact galaxies as stars causes the LF to be
systematically underestimated at the faint end by $\Delta \alpha =
0.03^{+0.02}_{-0.01}$, while the bright end remains unaffected. The
faint-end bias is comparable in size to the poisson errors of the MGC
and misclassified galaxies contribute $\sim$$6$ per cent to the LF in
the range $-17 < M_B < -14$~mag. In contrast, the luminosity density
is only affected at the $\sim$$2$ per cent level.

Clearly, we have {\em not} uncovered a large population of galaxies
`hiding' among the stars. On the other hand, high-SB selection effects
are not completely negligible either and hence should be taken into
account when constructing LFs. This is `naturally' achieved by first
constructing the bivariate space density of galaxies in either the
luminosity--SB or luminosity--size plane using a bivariate SWML
method, and then integrating over the second parameter to obtain the
LF \citep{Driver05}. Here we have confirmed that this method recovers
a result very close to an incompleteness corrected LF.

Our result regarding the {\em high} SB end of the galaxy distribution
should be viewed together with the situation at the {\em low} SB
end. While {\em luminous} galaxies are not affected by low SB
selection effects \citep[e.g.][]{Driver05}, {\em faint} galaxies are.
\citet{Blanton05} recently presented an SDSS LF corrected for low SB
incompleteness at $\mu_{{\rm eff},r} \ga 23$\mpass\ and concluded that
the correction resulted in a significant steepening of the faint end
of the LF. Indeed, at $M_B \ga -16$~mag the galaxy population extends
to even the deeper MGC's SB limit of $\mu_{\rm lim} = 25.25$\mpass\
\citep{Driver05}. Hence, even in the MGC the faint end of the LF is
significantly affected by both low and high SB selection effects,
despite the fact that the MGC is both deeper and of higher resolution
than any previous major ground-based survey. Clearly, to obtain a
complete view of the dwarf galaxy population requires still deeper and
higher resolution imaging (soon to be delivered by the VLT Survey
Telescope) as well as spectroscopy on 8-m class telescopes.



\section*{Acknowledgments}
The Millennium Galaxy Catalogue consists of imaging data from the
Isaac Newton Telescope and spectroscopic data from the Anglo
Australian Telescope, the ANU 2.3m, the ESO New Technology Telescope
(73.A-0092), the Telescopio Nazionale Galileo, and the Gemini
Telescope. The survey has been supported through grants from the
Particle Physics and Astronomy Research Council (UK) and the
Australian Research Council (AUS). The data and data products are
publicly available from http://www.eso.org/$\sim$jliske/mgc/ or on
request from JL or SPD.

Funding for the creation and distribution of the SDSS Archive has been
provided by the Alfred P.~Sloan Foundation, the Participating
Institutions, the National Aeronautics and Space Administration, the
National Science Foundation, the U.S.~Department of Energy, the
Japanese Monbukagakusho, and the Max Planck Society. The SDSS Web site
is http://www.sdss.org/. The SDSS is managed by the Astrophysical
Research Consortium (ARC) for the Participating Institutions. The
Participating Institutions are The University of Chicago, Fermilab,
the Institute for Advanced Study, the Japan Participation Group, The
Johns Hopkins University, the Korean Scientist Group, Los Alamos
National Laboratory, the Max-Planck-Institute for Astronomy (MPIA),
the Max-Planck-Institute for Astrophysics (MPA), New Mexico State
University, University of Pittsburgh, Princeton University, the United
States Naval Observatory, and the University of Washington.

\label{lastpage}

\end{document}